\documentclass[10pt]{iopart}
\usepackage{graphicx}
\usepackage{subfigure}
\usepackage{amsfonts}
\expandafter\let\csname equation*\endcsname\relax
\expandafter\let\csname endequation*\endcsname\relax
\usepackage{amsmath}
\usepackage{enumerate}
\usepackage{comment}
\usepackage{color}
\usepackage{citesort}

\newcommand*{\diff}{\mathop{}\!\mathrm{d}}

\DeclareMathOperator{\sgn}{sgn}
\newcommand{\calZ}{\mathcal{Z}}

\newcommand{\Order}[1]{\mathcal{O} ( #1 )}
\newcommand{\p}[1]{\left( #1 \right)}

\newcommand{\calN}{\mathcal{N}}

\newcommand{\bA}{\mathbf{A}}
\newcommand{\bpsi}{\boldsymbol{\psi}}
\newcommand{\calD}{\mathcal{D}}
\newcommand{\bx}{\boldsymbol{x}}

\newcommand{\hbx}{\hat{\boldsymbol{x}}}
\newcommand{\hx}{\hat{x}}
\newcommand{\hy}{\hat{y}}

\newcommand{\s}[1]{\left[ #1 \right]}
\newcommand{\calM}{\mathcal{M}}
\newcommand{\bomega}{\boldsymbol{\omega}}

\newcommand{\hP}{\hat{P}}
\newcommand{\bW}{\mathbf{W}}
\newcommand{\calA}{\mathcal{A}}
\newcommand{\expected}[1]{\left \langle #1 \right \rangle}
\newcommand{\bC}{\mathbf{C}}
\newcommand{\bL}{\mathbf{L}}

\newcommand{\bG}{\mathbf{G}}
\newcommand{\abs}[1]{\left| #1 \right|}

\newcommand{\balpha}{\boldsymbol{\alpha}}
\newcommand{\calP}{\mathcal{P}}
\newcommand{\Pop}{\textrm{pop}}

\begin{document}

\title{Heterogeneous mean-field analysis of the generalized Lotka-Volterra model on a network}

\author{Fabi\'an Aguirre-L\'opez}
\address{LadHyX UMR CNRS 7646, École polytechnique, 91128 Palaiseau Cedex, France}
\address{Chair of Econophysics and Complex Systems, École polytechnique, 91128 Palaiseau Cedex, France}

\ead{fabian.aguirre-lopez@ladhyx.polytechnique.fr}

\begin{abstract}
We study the dynamics of the generalized Lotka-Volterra model with a network structure. Performing a high connectivity expansion for graphs, we write down a mean-field dynamical theory that incorporates degree heterogeneity. This allows us to describe the fixed points of the model in terms of a few simple order parameters. We extend the analysis even for diverging abundances, using a mapping to the replicator model. With this we present a unified approach for both cooperative and competitive systems that display complementary behaviors. In particular we show the central role of an order parameter called the critical degree, $g_c$. In the competitive regime $g_c$ serves to distinguish high degree nodes that are more likely to go extinct, while in the cooperative regime it has the reverse role, it will determine the low degree nodes that tend to go relatively extinct.
\end{abstract}

\maketitle

% \tableofcontents

\section{Introduction}

Even though it was introduced almost 100 years ago, \cite{lotka1925elements,volterra1927variazioni}, the Lotka-Volterra model is still a topic of very active research in the field of dynamics of complex systems and theoretical ecology, \cite{akjouj2024complex}. 

For some years it has been recognized that ideas from theoretical ecology can be applied to finance, economy, and in general systems where agents can have interactions similar to ecological ones, \cite{scholl2021market,kamimura2011economic,Garnier-Brun_2021}. But even leaving the purely scientific curiosity aside, in recent years there has been a push for the use of agent based models (ABM's) in real economic scenarios \cite{farmer2009economy,naumann2023agent}. While this certainly has important advantages, it has at least one obvious issue. More complex models have lots of more parameters and lots of more possible behaviors. One possibility to overcome this problem is to explore systematically the possibly high-dimensional parameter space, which has been recently proposed in \cite{naumann2021exploration}. Alternatively, one can reduce the effective number of parameters by assuming that they are random numbers sampled from specified distributions. One can then study the behavior of the model varying the reduced number of degrees of freedom that parametrize the chosen distributions. The latter approach is very typical in the study of disordered systems from the statistical physics point of view, \cite{mezard1987spin}.

In recent years there has been extensive work in the intersection between statistical physics and theoretical ecology, precisely for the possibility of dealing with disordered systems with a large number of coupled variables. With the tools of statistical physics, the generalized Lotka-Volterra model has been widely explored and understood. Following the same approach that the one used for the replicator equations \cite{opper1992phase}, a close relative of the Lotka-Volterra model \cite{hofbauer1981occurrence}, a dynamical mean-field theory (DMFT) has been developed to explore many questions about the model, \cite{bunin2017ecological,galla2018dynamically}. This approach allows to study the Lotka-Volterra model and many extensions that approach more realistic settings. For this reason there has been an immense amount of work recently devoted to it, see \cite{biroli2018marginally,ros2022generalized,ros2023generalized,altieri2019constraint,altieri2021properties,de2024many,marcus2022local,lorenzana2022well,lorenzana2024interactions} and references within. The availability of these tools motivates us to study the Lotka-Volterra model as a minimal instance of a complex ABM, since the idea of competing agents is also natural in economy. 

In particular, we wish to explore the effect of having a network structure in a large complex system of interacting agents, with a special emphasis on the role of heterogeneity in the degrees of each node (the number of neighbors in the network). Large heterogeneous networks are well-known to mediate a large amount of human interactions, \cite{newman2018networks}. As an example from the area of economy, for the case of firm-level production networks, \cite{bacilieri2023firm}, it is well established that the underlying structures are highly heterogeneous with power law degree distributions. We find this to be a strong motivation to understand the effect of including a network structure in Lotka-Volterra type models.

From a theoretical point of view, dealing with a dynamical theory for disordered systems on an sparse graph is complicated endeavor \cite{mozeika2008dynamical,mozeika2009dynamical,mimura2009parallel,barthel2018matrix,machado2023improved,aguirre2022satisfiability,aurell2023closure,marcus2023local}. The amount of theory is much more limited compared to the one of the fully-connected counterpart. For the particular case of Lotka-Volterra, recent results show that indeed the behavior can be extremely complex even for networks with no degree heterogeneity. Therefore it is necessary to study the model in a regime where certain approximations can be made. Specifically one can focus on the high-connectivity regime. In this case, only a local quantity, the degree of each node will play an important role. While a lot {of} the complex discrete nature of the problem is lost, {many} interesting properties remain.

There is extensive literature of this important approximation in the field of mathematical modeling of epidemic spreading, \cite{pastor2001epidemic,pastor2015epidemic}. It is referred to as the degree based mean-field theory or heterogeneouse mean-field theory (DBMFT or HMFT).  In this area, this approximation has been successfully applied to deliver many of the standard results known and recently popularized during the COVID pandemic. Therefore it is natural to be hopeful that it can be equally useful for a qualitative understanding of large economic and ecological models. Indeed, recently these ideas have been applied in the field of statistical mechanics of disordered systems and of random matrix theory, \cite{metz2022mean,metz2020spectral,metz2023Ising,baron2022eigenvalue,park2024incorporating}. In this context, one needs to account not only for the heterogeneity of the degree of each node, but also for the heterogeneity in the strengths of the interactions. 

In this paper we develop a heterogeneous dynamical mean-field theory (HDMFT) for the generalized Lotka-Volterra model with network structure. In this way, one can take care of both types of heterogeneity simultaneously. We focus mainly on the regime where there is  typically a unique fixed point that the model can reach. Our main interest is to understand how the heterogeneity of the underlying degree distribution impacts the heterogeneity of the fixed point. While we maintain the language of theoretical ecology, speaking about species and abundances, we focus more on observables with an economic inspiration. That is, the Gini coefficient of the whole distribution of abundances, and the fraction of nodes that concentrate a finite fraction of the total abundance asymptotically. In order to present a unified framework for both competitive and cooperative systems, we also explore the diverging regime in which the Lotka-Volterra model is better described as a replicator model, \cite{hofbauer1981occurrence}.

This paper is organized as follows. In section \ref{sec:model} we define the model. In section \ref{sec:dynamics} we present the HDMFT. In section \ref{sec:pure} we present the solution for {the case of homogeneous interactions}. Building from the intuition of the homogeneous case, in section \ref{sec:general} we develop the theory for the fixed point in the general heterogeneous case. In section \ref{sec:linear} we show a simple linear stability analyisis of the theory, and in section \ref{sec:beyond-mean-field} we discuss the regime of validity of the mean-field assumption. Finally we discuss results and future directions in section \ref{sec:discussion}.

\section{The model \label{sec:model}}
We will focus on a generalization of the Lotka-Volterra model describing $N$ interacting species. The \emph{abundance} of each species is denoted by $x_i$, where $i = 1, \dots, N$. These quantities evolve according to the {following} differential equation, 
\begin{equation}
    \dot{x}_i(t) = x_i(t) \big( 1 - x_i(t) + \sum_j A_{ij} \alpha_{ij} x_j(t)\big)
    \label{eq:model-full}
\end{equation}
where we $A_{ij} \in \{0,1\}$ corresponds to an element of the adjacency matrix $\bA$ of the undirected interaction network and $\alpha_{ij} \in \mathbb{R}$ to the strength of an existing interaction. We take the adjacency matrix $\bA$ to be that of a random network sampled from a configuration model, {\cite{bollobas1980probabilistic}}, where the degree of each node, $k_i(\bA) = \sum_j A_{ij}$, is fixed to value $k_i$. The list of degrees, $\{k_i\}_{i=1,\dots,N}$ is a sample from a particular target distribution $p(k)$. We denote by $C$ the average degree, given by $C = \sum_k p(k) k$. We can write such distribution over \emph{symmetric} $\bA$'s using Kronecker deltas as
\begin{equation}
    p(\bA) = \frac{1}{\mathcal{Z}} \prod_{i=1}^N \delta_{k_i, \sum_j A_{ij}} .
    \label{eq:graph-ensemble}
\end{equation}
{where $\mathcal{Z}$ corresponds to the total number of networks with a given list of degrees.}

We introduce asymmetry in the interaction by taking the pairs $(\alpha_{ij}, \alpha_{ji})$ as i.i.d. random variables with
\begin{equation}
    \alpha_{ij} = \frac{\mu}{C} + \frac{\sigma}{\sqrt{C}} z_{ij}
\end{equation}
where {$(z_{ij},z_{ji})$ is a pair of jointly distributed Gaussian random variables with moments} $\overline{z_{ij}} = 0$, $\overline{z_{ij}^2} = \overline{z_{ji}^2} = 1$, and $\overline{z_{ij}z_{ji}} = \gamma$. In this way the correlation between the weights is controlled by $\gamma$, this changes the proportion of the different type of interactions that can occur between species, see \cite{galla2018dynamically}. We will always consider that the model starts from an initial condition $\bx_0$ whose entries are random i.i.d. numbers sampled from a given distribution $p(x_0)$.

In Section \ref{sec:pure} we will specialize to the homogenuous case, which corresponds to setting $\sigma = 0$. This means all interaction have the same value,
\begin{equation}
    \dot{x}_i(t) = x_i(t) \big( 1 - x_i(t) + \frac{\mu}{C} \sum_j A_{ij} x_j(t)\big).
    \label{eq:model-competition}
\end{equation}
The only disorder to consider is that of the random matrix $\bA$. In this case if we have $\mu > 0$, it corresponds to cooperation between all species and if we have $\mu<0$ it corresponds to pure competition between everyone.

\section{Dynamical theory \label{sec:dynamics}}

Our aim is to give a statistical description of the solutions of \eqref{eq:model-full} with tools from statistical physics. The objective is to write down a theory that will predict averages of $p$-point functions of single species over the population for a \emph{single} instance of $\balpha$ and from a \emph{single} initial condition $\bx_0$,
\begin{equation}
    \expected{f}_{\Pop} = \frac{1}{N}\sum_{i=1}^N f\big( x_i(t_1; \bx_0, \bA, \balpha ),x_i(t_2; \bx_0, \bA, \balpha ),\dots,x_i(t_p; \bx_0, \bA, \balpha )\big).
\end{equation}
where $x_i(t; \bx_0, \bA, \balpha )$ denotes a solution of \eqref{eq:model-competition} for a given initial condition $\bx_0$, given interaction matrix $\bA$, and given interaction strengths $\balpha$.
Examples could be the first and second moments,
\begin{align}
    \expected{x(t)}_{\Pop} & = \frac{1}{N} \sum_{i=1}^N x_i(t; \bx_0, \bA, \balpha) \\
    \expected{x(t)x(t')}_{\Pop} & = \frac{1}{N} \sum_{i=1}^N x_i(t; \bx_0, \bA, \balpha) x_i(t'; \bx_0, \bA, \balpha)
\end{align}
or even the full empirical distribution of $x_i$'s ,
\begin{equation}
    Q_N(x|t) = \expected{\delta(x - x(t))}_{\Pop} = \frac{1}{N} \sum_{i=1}^N \delta\big(x - x_i(t; \bx_0, \bA, \balpha)\big).
\end{equation}

In order to develop a theory, we need to explore the asymptotic limit of large $N$ and large $C$, particularly the case when $N \gg C \gg 1$. More precisely, it means considering degree distribution where $C$ is growing sublinearly with $N$, for example $C = \Order{\log N}$. This is equivalent to analytically first taking $N\to \infty$ and then taking $C\to \infty$ in the asymptotic theory in $N$.  We show that in this regime, population averages converge to non fluctuating quantities that depend only on the distributions associated with $\bx_0$, $\bA$, and $\balpha$. That is, the theory should only depend on the distribution of $\alpha_{ij}$'s, determined by $\mu$, $\sigma$, and $\gamma$, the distribution of degrees, $p(k)$, and the distribution of the initial condition, $p(x_0)$. Even though we develop a theory in this asymptotic limit, we will show that it is in very good agreement with results from simulations done with relatively small system sizes, $N\sim 10^3$, $C\sim 50$.

Following the approach used in \cite{metz2023Ising} for spin systems, in \ref{app:generating-functional} we show the populations averages will converge to averages over an \emph{effective} single stochastic process $x$, defined by 
\begin{equation}
    \label{eq:effective-process}
    \dot x = x \p{1 - x + g \mu M(t) + \sqrt{g} \sigma \eta(t) +  g \gamma \sigma^2 \int_0^t \diff s\, G(t,s) x(s)},
\end{equation}
where $g$ is a random positive number, $\eta(t)$ is random correlated zero average Gaussian noise, and the functions $M(t)$ and $G(t,s)$ are fixed by the theory in the way detailed below. We also assume that the initial condition is random and sampled from $p(x_0)$.

Population averages can then be substituted by those over the stochastic process, 
\begin{equation}
    \expected{f}_{\textrm{pop}} \underset{N\to \infty}{\longrightarrow} \expected{f}_{\textrm{eff}}
\end{equation}
where $\expected{\circ}_{\textrm{eff}}$ means averaging over the effective process \eqref{eq:effective-process}. For example, for a one-point function we would have, 
\begin{equation}
    \expected{f(x(t))}_{\textrm{eff}} = \int \diff x_0 p (x_0) \diff g \nu(g) \expected{f(x(t;x_0,g,\eta))}_{\eta}
\end{equation}
where $x(t;x_0,g,\eta)$ means a solution of \eqref{eq:effective-process} and $\expected{\circ}_\eta$ denotes averaging over the distribution of the noise $\eta(t)$.

We still need to define the distribution for $g$, denoted as $\nu(g)$, and the correlation function for the noise $\eta(t)$, denoted as
\begin{equation}
    \expected{\eta(t)\eta(s)}_{\eta} = C(t,s).
\end{equation}
Therefore in order to fully characterize \eqref{eq:effective-process} we need to specify how to define $\nu(g)$, $M(t)$, $C(t,t')$, and $G(t,s)$.

Since we are looking at the high connectivity limit, $\nu(g)$ is defined as the high connectivity limit of the degree distribution, that is
\begin{equation}
    \nu(g) = \lim_{C\to \infty} \sum_{k=0}^\infty p(k) \delta\hspace{-1mm}\p{g - \frac{k}{C}}
    \label{eq:rescaled-degrees}
\end{equation}
even though $g$ is a continuous variable, we will speak interchangeably both of $g$ and of $k$ as of the degree. To return {to} the language of $k$ one need only to multiply by $C$ in {any} finite instance. 

The functions $M(t),C(t,s),G(t,s)$ turn out to be moments over an associated but different stochastic process. We will refer to it as the cavity effective process, in reference to the cavity or belief propagation method, \cite{mezard2009information}. The only difference with \eqref{eq:effective-process} is the distribution for $g$, in this case it is given by the cavity distribution
\begin{equation}
    \label{eq:cavity-distribution}
    g \sim \nu_{\textrm{cav}}(g) = g \nu(g)
\end{equation}
If we define $\expected{\circ}_{\textrm{cav}}$ as averages over \eqref{eq:effective-process} but where $g$ is sampled from $\nu_{\textrm{cav}}$ instead of $\nu(g)$, then we can write down the defining self-consistency relations for $M(t)$, $C(t,t')$, and $G(t,t')$;
\begin{align}
    M(t) & = \expected{x(t)}_{\textrm{cav}}, \label{eq:M-cavity}\\
    C(t,t') & = \expected{x(t)x(t')}_{\textrm{cav}}, \\
    G(t,t') & = \expected{\frac{1}{\sqrt{g} \sigma }\frac{\delta x(t)}{\delta \eta(t')}}_{\textrm{cav}}.\label{eq:G-cavity}
\end{align}

We will analyze two cases, one in which equilibrium is used, for small enough $\mu$,
\begin{equation}
    \lim_{t\to\infty} M(t) = M^*
\end{equation}
and another in which the average diverges in a finite time $t^*$, 
\begin{equation}
    \lim_{t \to t^*} M(t) = \infty
\end{equation}
for high enough $\mu$. While this regime is typically not considered, since it has no meaning in the ecological context, we will allow ourselves to explore it. In this way we present a full picture for the cooperative case of $\mu>0$ in more detail.

To study the diverging case we need to rescale the abundance, $x$, in order to find a stable distribution as the average diverges, 
\begin{equation}
    y(t) = \frac{x(t)}{\mu M(t)},
\end{equation}
and a new time scale defined by
\begin{equation}
    \tau = \mu \int_0^t \diff s \, M(s).
\end{equation}

In this new time variable, $\tau$, there is a well defined $\tau\to\infty$ and there is no more divergence in finite time. With the abundance rescaling we also expect finite distributions.

We can then look at the effective dynamics for this new variable,
\begin{align}
    \dot y = {} & y \p{ - y + g + \sqrt{g} \sigma \xi(\tau)  + g \gamma \sigma^2  \int_0^\tau \diff \zeta \, H(\tau, \zeta) y(\zeta)  - \mu B(\tau)} 
\end{align}
where $\xi(\tau)$ is Gaussian noise with correlations given by 
\begin{equation}
    \expected{\xi(\tau)\xi(\zeta)}_{\xi} = q(\tau,\zeta), 
\end{equation}
and where we have rescaled the response kernel $G$ in the following way,
\begin{equation}
    H(\tau,\zeta) = G(\tau,\zeta)\frac{M(\zeta)}{M(\tau)}.
\end{equation}
Again, we have the following self-consistent set of equations,
\begin{align}
    \label{eq:B-tau}
    1 = {} & \mu\expected{y(\tau)}_{\textrm{cav}}\\
    q(\tau,\zeta) = {} & \expected{y(\tau)y(\zeta)}_{\textrm{cav}} \\
    H(\tau,\zeta) = {} & \expected{\frac{1}{\sqrt{g}\sigma} \frac{\delta y(\tau)}{\delta \xi(\zeta)}}_{\textrm{cav}}
\end{align}
We have arrived {to a result} that looks similar to a theory of the replicator model like the one developed in \cite{opper1992phase}, but one that includes degree heterogeneity and that fixes the cavity average and not the true average.  By including the analysis of the model in this regime, we will show how certain observables of the model \eqref{eq:model-full} vary smoothly as $\mu$ is varied in the whole range $(-\infty,\infty)$, even if $\mu_c$ is crossed into the diverging regime.

\subsection{Annealed approximation}

As mentioned in the introduction, the resulting theory is an example of the heterogeneous mean-field type \cite{pastor2001epidemic}, but adapted to this type of process and with the self consistent kernels to take into account the complexity of the mean-field. As a matter of fact, within this theory in the high connectivity case, much of the original structure of the graph $\bA$ is not relevant, since it can be derived by making the so called annealed graph approximation, \cite{carro2016noisy}. It consists of substituting the random graph by a fixed fully connected matrix $\Tilde{\bA}$, by the following rank one matrix
\begin{equation}
    A_{ij} \rightarrow \Tilde{A}_{ij} = \frac{k_i k_j}{NC}.
\end{equation}
While the theory can be derived starting from this assumption, in our case in \ref{app:generating-functional} we derive \eqref{eq:effective-process} starting from the assumption there is indeed an underlying sparse graph and obtain these results as an expansion in $1/C$. This allows us also to understand when the theory should not hold.

This type of approximation was also discussed in \cite{Garnier-Brun_2021} to capture main features of correlation matrices. Indeed similar results to those of Section \ref{sec:pure} were found.

\section{The homogeneous case\label{sec:pure}}

It is particularly interesting to look at the model without noise, that is $\gamma = \sigma^2 = 0$, same setting as the one studied in \cite{marcus2022local}. In this case our theory is greatly simplified since a lot of the disorder has been removed. The HDMFT, \eqref{eq:effective-process}, turns into a set of equations that are actually solvable with the help of a computer. Even more, it is also an interesting minimal model of cooperation/competition in the presence of heterogeneity, \cite{Garnier-Brun_2021}.
In this case the dynamics, \eqref{eq:effective-process}, reduces to,
\begin{equation}
    \dot x = x \big( 1 - x + g \mu M(t)\big).
    \label{eq:cavity-uniform}
\end{equation}

The differential equation, \eqref{eq:cavity-uniform}, can actually be solved explicitly,
\begin{equation}
    x(t;x_0,g,M) = \frac{x_0 \exp{\int_0^t \diff s \, (1 + g \mu M(s))}}{ 1 + x_0 \int_0^t \diff s \, \exp\p{\int_0^s \diff v
    (1 + g \mu M(v)) }}.
    \label{eq:solution-competitive}
\end{equation}
The only unknown function is the cavity average, $M(t)$, for which we can write the following equation, derived from \eqref{eq:M-cavity},
\begin{equation}
    M(t) = \int \diff x_0 \, p(x_0) \diff g \, \nu(g) \, g \, x(t;x_0,g,M).
    \label{eq:cavity-M-competitive}
\end{equation}
Once this equation is solved, we can actually solve the dynamics of the distribution of abundances at all times, which an be easily shown to be
\begin{equation}
    \begin{aligned}
        Q(x|t) & =  \int \rmd g \, \nu(g) \, P_g(x|t), \\
        P_g(x|t) & = \int \rmd x_0 \, p(x_0) \, \delta\big(x - x(t;x_0,g,M)\big),
    \end{aligned}
\end{equation}
where we have introduced the full distribtuion of abundances at time $t$, denoted as $Q(x|t)$, and the conditional distribution on degree $g$, $P_g(x|t)$, telling us the distribution of abundances for all nodes with the same degree. 

We can also write a simple equation with a closed form solution in the diverging regime,
\begin{equation}
    \dot y  = y ( -y + g - \mu B(\tau))
\end{equation}
with solutions analogous to the other, where $B(\tau)$ satisfies the self-consistent relation,
\begin{align}
    1 = \mu \expected{y(\tau;y_0,g,B)}_{\textrm{cav}}.
\end{align}

\begin{figure*}
    \centering
    \begin{picture}(370,125)
    % \put(0,0){\line(1,0){370}}
    % \put(0,0){\line(0,1){125}}
    \put(10,10){\includegraphics[trim = 28 14 0 0 ,clip,width=.49\textwidth]{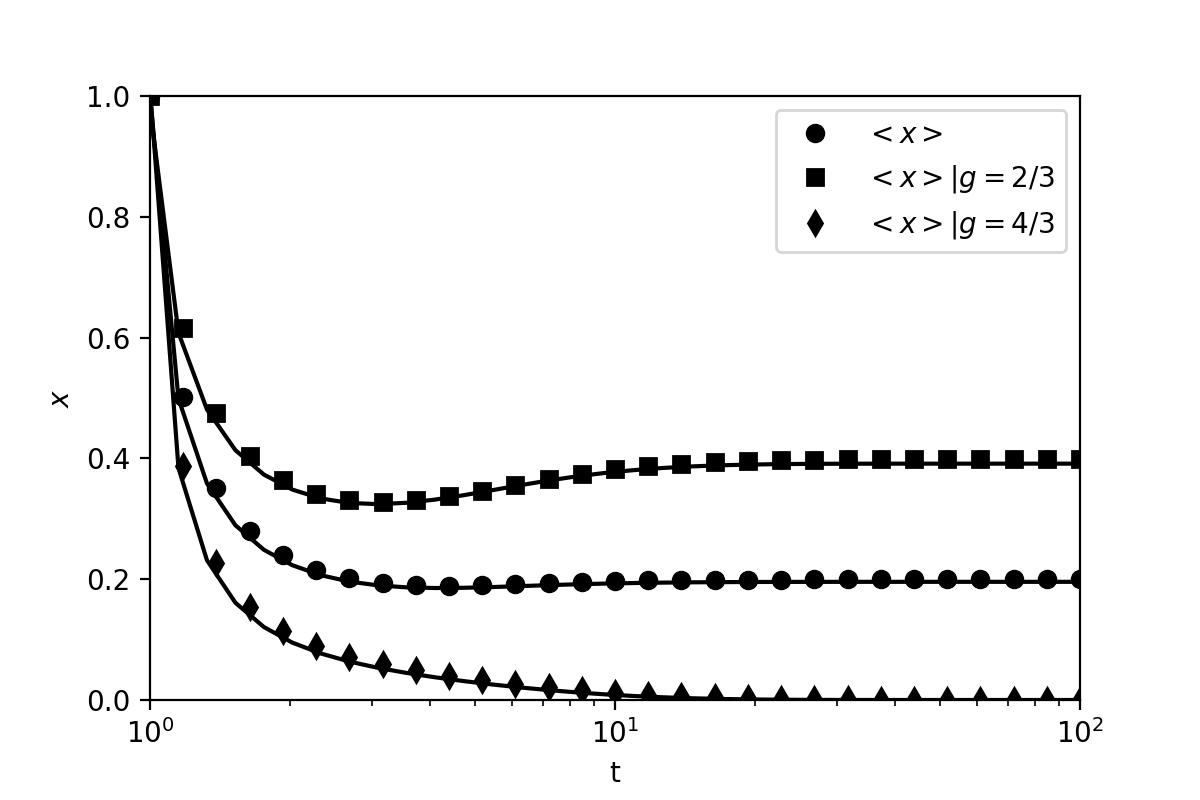}}
    \put(200,10){\includegraphics[trim = 25 14 0 0 ,clip,width=.49\textwidth]{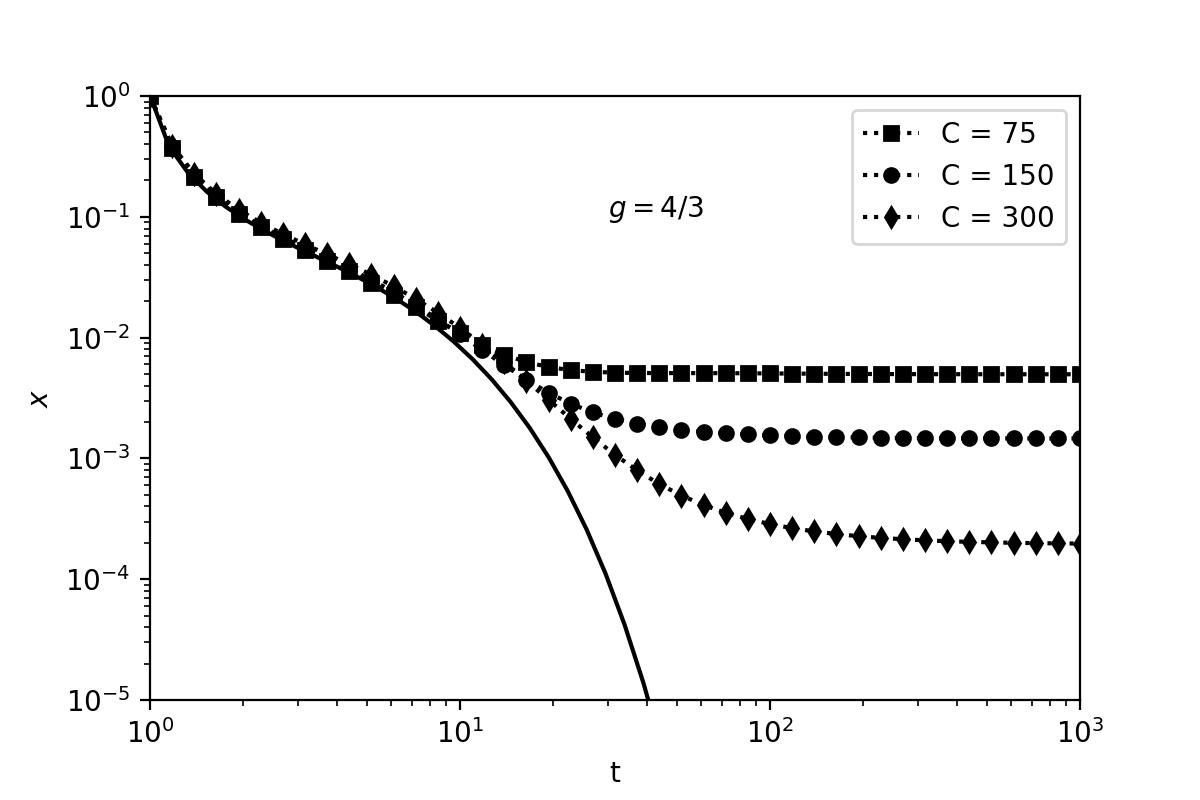}}
    \put(95,0){\scalebox{1.2}{$t$}}
    \put(290,0){\scalebox{1.2}{$t$}}
    \put(0,70){\scalebox{1.2}{$x$}}
    \put(187,70){\scalebox{1.2}{$x$}}
    \end{picture}
    \caption{Left: Average abundance,{ $\expected{x} = \int \diff x Q(x|t) x$, and conditional one, $\expected{x}_{|g}  =\int \diff x P_g(x|t) x$}, as a function of time for bimodal graph \eqref{eq:bimodal-distribution}, with $N = 3000$, $C = 300$, {and $\mu = -7$}. Right: $P_g(x|t)$ for $g = 4/3$ for different values of $C$. Symbols represent numerical {integration of a single instance of} \eqref{eq:model-competition} and solid lines of the solution of the theory \eqref{eq:cavity-M-competitive}.}
    \label{fig:dynamics}
\end{figure*}

Amazingly, the solution of the theory can describe well the statistics of a single instance. To show this, we compare with numerical results for a single random graph with a bimodal degree distribution, in this case where half {of the nodes of the graph have degree}  $2C/3$ and the other half $4C/3$. To get numerical results we sampled a single graph with $N=3000$ and degrees $200$ and $400$, $C=300$, and then integrated \eqref{eq:model-competition} numerically. For the theoretical results it is necessary to calculate the rescaled degree distribution \eqref{eq:rescaled-degrees}, which gives
\begin{equation}
    \nu(g) = \frac{1}{2}\delta\p{g - \frac{2}{3}} + \frac{1}{2}\delta\p{g - \frac{4}{3}}.
    \label{eq:bimodal-distribution}
\end{equation}
With this we can calculate the distribution of abundances over time, in particular compare the total average, and the degree conditional averages, $\int \rmd x \, P_g (x|t) x$. In the left panel of Figure \ref{fig:dynamics} we see an almost perfect match as the system approaches the fixed point. In the right panel of Figure \ref{fig:dynamics} we show a finite $C$ effect. While the asymptotic theory predicts extinction of species with $g = 4/3$, which means an exponential vanishing over time, in the finite $C$ case we observes it reaches a plateau that vanishes when $C\to\infty$.

In the next section we will expand on the properties of the fixed point.

\subsection{The fixed point}

Once the system evolves for a long time, one expects it to approach a fixed point. By looking at the long time behaviour of the theory \eqref{eq:cavity-uniform}, we can show it is consistent with an approach to equilibrium. In section \ref{sec:beyond-mean-field} we will discuss the differences at finite $C$ where the picture is completely different. 

To characterize the fixed point we take the limit $t \to \infty$. If we assume $\lim_{t \to \infty} M(t) = M^*$, we can derive from the solution of \eqref{eq:cavity-uniform} the statistics of the fixed point, \cite{opper1992phase,galla2018dynamically}. First, to derive an equation for $M^*$ one should observe that there should exist at time $T$ such that $M(t) \approx M^* $ for $t>T$, which means that if we define $x_T = x(T;x_0,g,M)$, we can write the solutions at time $t$ as 
\begin{equation}
    \begin{aligned}
        x(t;x_0,g,M) & \approx \frac{x_T \exp\p{(t- T) (1 + g \mu M^*)}}{1 + x_T \displaystyle\frac{\exp\p{(t- T) (1 + g \mu M^*)}}{1 + g \mu M^* }}
    \end{aligned}
\end{equation}
which implies that 
\begin{equation}
    \lim_{t \to \infty} x(t;x_0,g,M) = \left\lbrace \begin{array}{cc}
    1 + g \mu M^*     & \textrm{if } 1 + g \mu M^* >0 \\
    0    & \textrm{if } 1 + g \mu M^* \le 0
    \end{array}\right. 
    \label{eq:long-time-limit-xt-minimal}
\end{equation}
This means that the theory predicts extinctions for certain species in the case of competition, $\mu<0$, since the abundance will vanish if the degree $g$ is large enough. To survive, the degree of a species should not exceed the critical value given by
\begin{equation}
\label{eq:g_c}
    g_c = - \frac{1}{\mu M^*}
\end{equation}

We can see that in the homogeneous case {everything} is determined by the degree, $g$, so the conditional distribution becomes concentrated in delta functions,
\begin{equation}
    P_g(x^*) = \delta(x^* - 1 - g\mu M^*)\theta(1 + g \mu M^*) + \delta(x^*) \theta(-1-g\mu M^*)
    \label{eq:Pg-homogeneous}
\end{equation}
The cavity equation, \eqref{eq:cavity-M-competitive}, at equilibrium becomes
\begin{equation}
    M^* = \frac{\expected{g\,\theta(1 + g\mu M^*)}}{1 - \mu \expected{g^2\,\theta(1 + g\mu M^*)}},
    \label{eq:M-star}
\end{equation}
where $\expected{f(g)}$ represents an average over the rescaled degree distribution, $\nu(g)$.

We point out the two important properties about \eqref{eq:M-star}. For $\mu<0$ it is a self-consistent equation, since we are using averages only over surviving species, those ones with degree $g < g_c$ and $g_c$ depends on $M^*$ by \eqref{eq:g_c}.  In the cooperative case, $\mu>0$, there exists a critical value of interaction strength{, $\mu_c$}, above which $M^*$ diverges meaning that there is no equilibrium. {By inspection of \eqref{eq:M-star}, we can deduce it should be given by
\begin{equation}
    \mu_c = \frac{1}{\expected{g^2}}.
\end{equation}
}
This is the same the same regime as the one discussed in the introduction, therefore $\mu>\mu_c$ determines the beginning of the diverging regime in this case.

In the case where $M^*$ exists we can write down the 
equilibrium distribution $Q(x^*)$. We notice that it can all be written in terms of the critical degree $g_c$, even for $\mu>0$,
\begin{equation}
    Q(x^*) = \abs{g_c}\nu\p{- g_c \p{x^* - 1}} \theta(x^*) + \theta(-\mu)\delta(x^*) \int_{g_c}^\infty \rmd g \, \nu (g) 
    \label{eq:equilibrium-minimal-model}
\end{equation}
This expression is valid also for $0<\mu<\mu_c$ since following \eqref{eq:g_c} and \eqref{eq:cavity-M-competitive}, even though it is negative, $g_c$ is also well defined in this range. This result , \eqref{eq:equilibrium-minimal-model}, has a very clear physical meaning, at the mean-field level the abundance at the fixed point is only determined by the number of neighbors of each node, the degree $g_i$, whose fraction is in turn determined by the rescaled degree distribution $\nu(g)$. The first term 
 in \eqref{eq:equilibrium-minimal-model} is simply a rescaling of $\nu(g)$ (reversed for $\mu<0$), and the second term only exists for $\mu<0$ since it corresponds to the fraction of extinct nodes. Because degrees are always positive, this implies that $\nu(g) \propto \theta(g)$, which then implies the following,
\begin{equation}
    \begin{aligned}
        \mu>0 \Rightarrow x^*\ge 1\\
        \mu<0 \Rightarrow x^*\le 1
    \end{aligned}
\end{equation}

\begin{figure*}
    \centering
    \begin{picture}(370,125)
    % \put(0,0){\line(1,0){370}}
    % \put(0,0){\line(0,1){125}}
    \put(10,10){\includegraphics[trim = 35 20 0 0 ,clip,width=.49\textwidth]{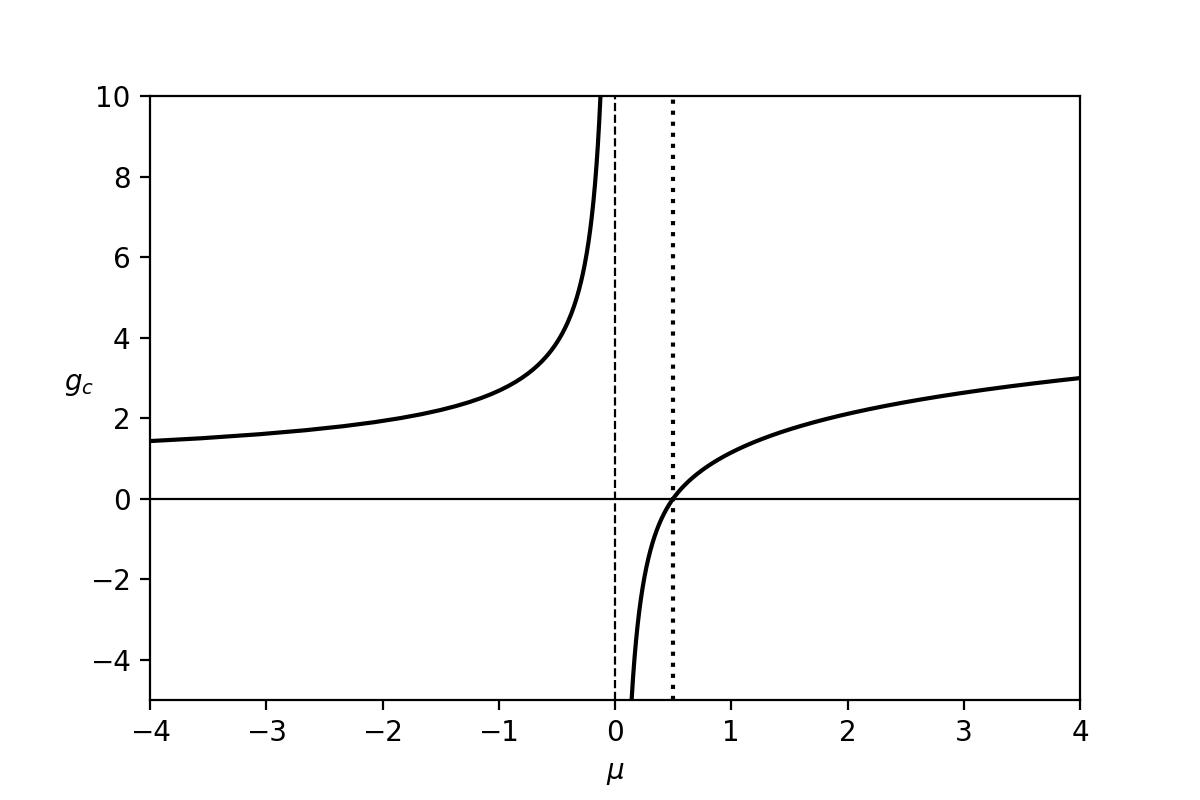}}
    \put(205,10){\includegraphics[trim = 26 15 0 0 ,clip,width=.49\textwidth]{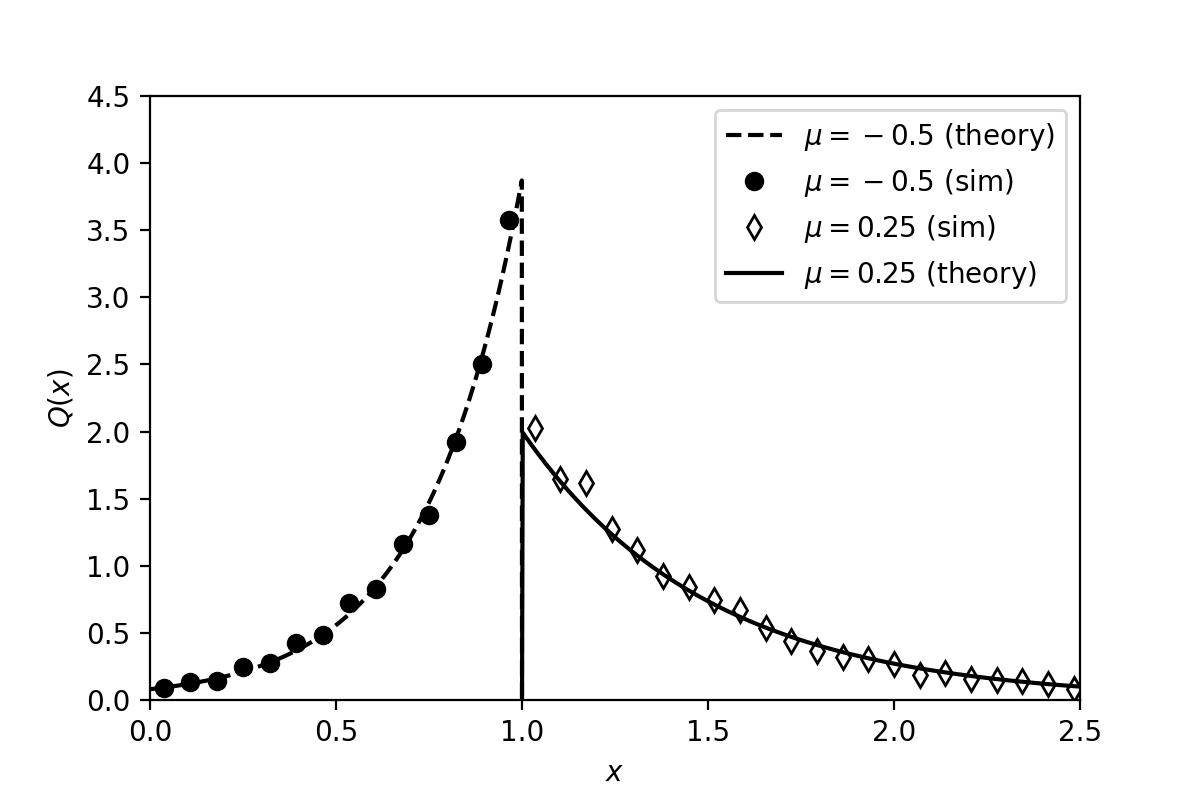}}
    \put(93,0){\scalebox{1.2}{$\mu$}}
    \put(290,0){\scalebox{1.2}{$x^*$}}
    \put(0,70){\scalebox{1.2}{$g_c$}}
    \put(180,70){\scalebox{1}{$Q(x^*)$}}
    \end{picture}
    \caption{Left: Solution for the critical degree $g_c$ from \eqref{eq:g_c-equation} for exponential degree distribution, $\nu(g)=\rme^{-g}$. Right: Abundance distribution at the fixed point for homogeneous model, \eqref{eq:model-competition}, with $N=3000$ and $C=50$ for $\mu = -0.5$ (competitive) and $\mu = 0.25$ (cooperative). Symbols are  averages over 50 instances of numerical simulations. Dashed and solid lines correspond to the theory \eqref{eq:equilibrium-minimal-model}. {Error bars are of the order of magnitude of the symbols and omitted for clarity.}}
    \label{fig:equilibrim_and_survivors}
\end{figure*}
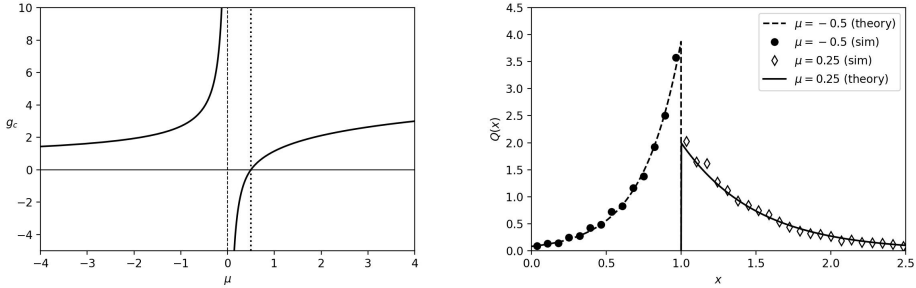

If we explore the diverging case, when $\mu>\mu_c$, as expected there is no solution for $M^*$ because the average abundance diverges. But, we see that $y$ also approaches a fixed point in a similar way. If we assume that $\lim_{\tau\to\infty} B(\tau) = B^*$, then we have that for large $\tau$,
\begin{equation}
    \dot y \approx y(g-y- \mu B^*),
\end{equation}
and we again see that the fixed point also depends only on the value of $g$,
\begin{equation}
    \lim_{\tau \to \infty} y(\tau) = \left\lbrace\begin{array}{cc}
        0 &  \textrm{if } g<\mu B^* \\ 
        g - \mu B^* & \textrm{if } g>\mu B^*
    \end{array}\right.
\end{equation}
where $B^*$ satisfies a similar equation to \eqref{eq:M-star} derived from \eqref{eq:B-tau}.
We see that there is a fraction of nodes that go extinct in the $y$ variable, meaning that even though their value of $x$ is diverging in the true variable, it is vanishing with respect to the others. Therefore, we can also define a critical value for $g$,
\begin{equation}
    g_c = \mu B^*,
\end{equation}
but in this case surviving nodes are the ones with degree \emph{above} $g_c$. From this it follows that the asymptotic distribution for $y^*$ is
\begin{equation}
    Q(y^*) = \nu(y^* + g_c)  + \delta(y^*) \int_{0}^{g_c} \diff g \, \nu(g).
    \label{eq:y-minimal-equilibrium}
\end{equation}

This actually gives a physical meaning of why $g_c$ is negative in the range $0<\mu<\mu_c$. In the cooperative case in general, nodes with higher degree are favored as they have more collaborators.  Therefore it is natural to assume the cut-off should be from below. For positive values of $\mu$ smaller than $\mu_c$ the critical value $g_c$ is negative because everyone survives. Once $\mu = \mu_c$ we have $g_c=0$ and then it becomes positive to give the fraction of extinct species with low degrees in \eqref{eq:y-minimal-equilibrium}.

As a matter of fact, we can write the whole theory in terms only of $g_c$, forgetting about $M^*$ and $B^*$, as it can be seen from \eqref{eq:equilibrium-minimal-model} and \eqref{eq:y-minimal-equilibrium}. It is enough to write down an self-consistent equation for $g_c$ valid in the whole range of $\mu$. Both equations for $M^*$ and $B^*$ turn into the same equation, provided we calculate the moments of the degree of surviving species in the correct way,
\begin{equation}
    g_c = \frac{\expected{g^2}_{g_c}}{\expected{g}_{g_c}} - \frac{1}{\mu}\frac{1}{ \expected{g}_{g_c}}.
    \label{eq:g_c-equation}
\end{equation}
where
\begin{equation}
    \expected{f(g)}_{g_c} = \left\lbrace \begin{array}{cc}
    \displaystyle\int_{0}^{g_c} \diff g \, \nu(g) \, f(g)     & \textrm{if } \mu < 0 \vspace{3mm}\\
    \displaystyle\int_{g_c}^\infty \diff g \, \nu(g) \, f(g)     & \textrm{if } \mu> 0
    \end{array}\right.
\end{equation}
This equation was previously derived in a similar context but with a completely different motivation in \cite{Garnier-Brun_2021}. We show in Figure \ref{fig:equilibrim_and_survivors} that $g_c$ is a well-defined function in the whole range of $\mu$ with a singularity only at $\mu = 0$, where there are no interactions in the homogeneous case.

A good agreement is found between theory and numerical simuations for the abundance distribution when the fixed point is reached. In the right panel of Figure \ref{fig:equilibrim_and_survivors} we compare $Q(x^*)$ as predicted by the theory and as observed in a sample of 50 simulations with $N = 3000$ and $C = 50$ for an exponential distribution, $\nu(g)$. We see a good match both in the competitive and cooperative regimes.

\subsection{The asymptotic inequality of the model}

Since we are interested in the inequality of the system we quantify it through the Gini coefficient, a standrad measure in economic literature. It quantifies the difference in area between the Lorenz curve of a perfectly equal distribution and a particular distribution of interest. In terms of the cumulative of the distribution of interest, $F(x)$,  it can be calculated by the formula
\begin{equation}
    G = 1 - \frac{1}{\expected{x}} \int_{0}^\infty \diff x \, \p{1 - F(x)}^2 .
    \label{eq:gini-formula}
\end{equation}
In particular we will apply it to $Q(x^*)$ and $Q(y^*)$.

By inspecting \eqref{eq:equilibrium-minimal-model} we can see that at $\mu = 0$ then $G =0 $ since all species relax to $x_i = 1$. This is the state of perfect equality. In the absence of heterogeneity, that is with $\nu(g) = \delta(g-1)$, we have equality for all values of $\mu$ since the equilibrium solution is $x^* = 1 / ( 1 - \mu)$ for all species. Nevertheless, for the heterogeneous case, $\expected{g^2}>{\expected{g}}^2$, the inequality will increase if in \emph{both} directions, that is favoring competition or cooperation. This is a significant effect of having a heterogeneous degree distribution. The inequality increases in the $\mu<0$ case as the fraction of extinct nodes increase. In the case of $\mu>0$, we see that in the range $\mu<\mu_c$ the Gini coefficient is increasing as species begin interacting. At the critical point it reaches the Gini coefficient of the degree distribution itself, Which can be seen by setting $g_c=0$ in \eqref{eq:y-minimal-equilibrium}. Afterwards, for $\mu>\mu_c$, the inequality goes \emph{over} the imposed one due to the relative extinction of the species with low degree.

\begin{figure*}
    \centering
    \begin{picture}(220,150)
    % \put(0,0){\line(1,0){220}}
    % \put(0,0){\line(0,1){150}}
    \put(12,10){\includegraphics[trim = 32 20 0 0 ,clip,width=.6\textwidth]{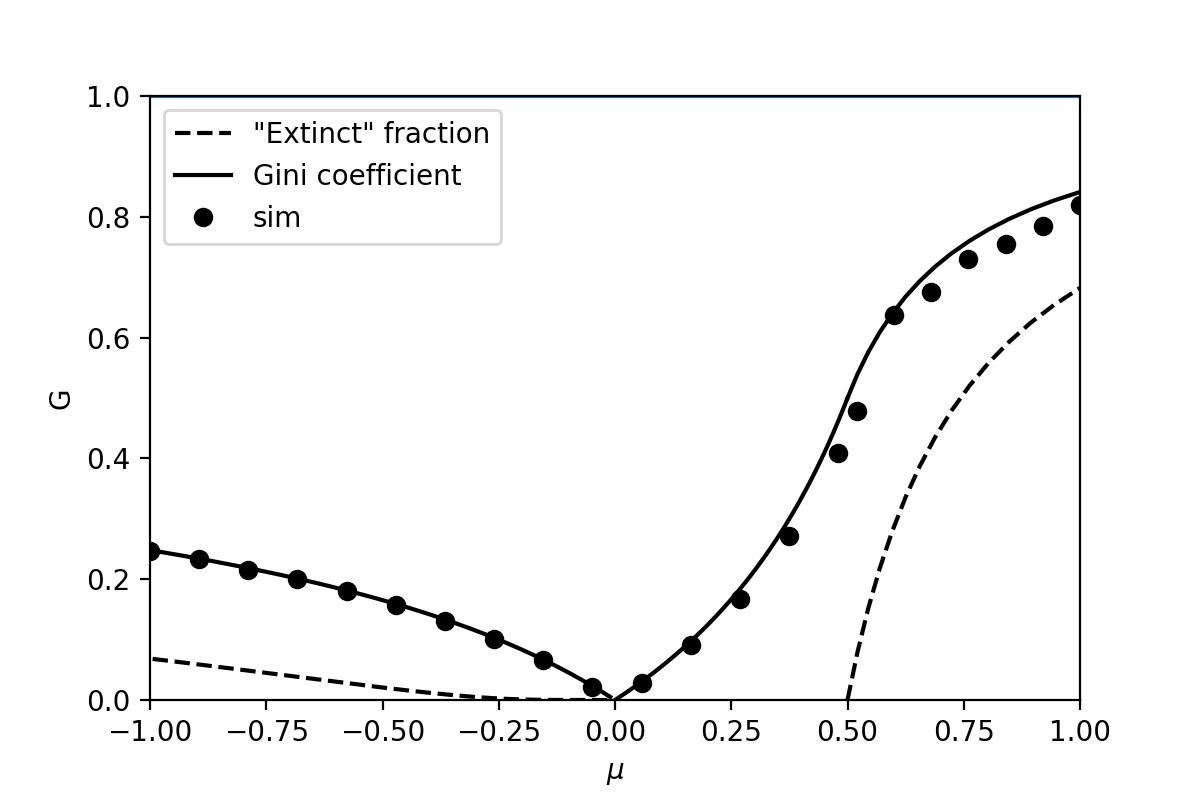}}
    \put(114,0){\scalebox{1.2}{$\mu$}}
    \put(0,75){\scalebox{1.2}{$G$}}
    \end{picture}
    \caption{Gini coefficient of the abundance distributions $Q(x^*)$ and $Q(y^*)$ at the fixed point for different values of $\mu$. The solid line corresponds to the Gini coefficient predicted by the theory, \eqref{eq:equilibrium-minimal-model} and \eqref{eq:y-minimal-equilibrium}, and the solid circles corresponds to the Gini coefficient measured directly for a single realization of the model, \eqref{eq:model-competition}, with $N = 3000$ and an exponential degree distribution with $C = 50$. Dashed lines corresponds to the fraction of extinct species. For values $\mu>1/2$, it corresponds to relative extinctions. It is interesting that the rise of the Gini coefficient is sharper in the cooperative side, $\mu>0$, than in the competitive one $\mu<0$.}
    \label{fig:gini-homogeneous}
\end{figure*}

For example, for an exponential degree distribution,  $\nu(g) = \rme^{-g}$, we can write a simple equation for $g_c$, and therefore an exact expression for the Gini coefficient can be derived from \eqref{eq:equilibrium-minimal-model} and \eqref{eq:y-minimal-equilibrium}. In Figure \ref{fig:gini-homogeneous} we show a very good match between theory and single instances of the model.

\section{Fixed point of the general case \label{sec:general}}

In the general case when $\sigma,\gamma \not = 0$, it is hard to say something about the dynamics at an arbitrary time $t$ from the general theory. Numerical solutions of \eqref{eq:effective-process}, even in the homogenous degree case, $\nu(g) = \delta(g-1)$, are computationally very expensive to solve, \cite{roy2019numerical,manacorda2020numerical}. Even more, with a non-trivial $\nu(g)$ the computation time would increase dramatically.

Nevertheless, it is possible to work out properties of the fixed point reached dynamically by doing similar considerations as in the previous section. We assume there exists a time $T\gg 1$, such that at this points the dynamics is very close to a fixed point $x^*$, and therefore the effective dynamics are
\begin{equation}
    \dot{x} \approx x(t) ( 1 - x(t) + g \mu M^*  + \sqrt{g} \sigma \eta^*  + \gamma g \sigma ^2 \chi^* x(t) ),
    \label{eq:fixed-point-approach-xt}
\end{equation}
where $\eta^* \sim P(\eta^*)$ is a fixed Gaussian number with variance $C^* = \expected{{x^*}^2}_{\textrm{cav}}$, which should be the long time behavior for $\eta(t)$ if there is a fixed point, and where we have replaced the response integral for its behavior at long times assuming $x(t)$ is close to its final value,
\begin{equation}
    \chi^* = \int_{0}^\infty \rmd s \, G(s).
    \label{eq:integrated-response}
\end{equation}
Following the same argument as for \eqref{eq:long-time-limit-xt-minimal}, we see that the value of $x^*$ depends on the relationship between the constant terms in the growth rate and the coefficients of $x(t)$ in \eqref{eq:fixed-point-approach-xt}. We give the result in terms of $P_{g,\eta^*}(x^*)$, the distribution of abundance $x^*$ at the fixed point given a value of $g$ and of $\eta^*$,
\begin{equation}
    \begin{aligned}
        P_{g,\eta^*}(x^*) & = \delta(x^* - f^*) \theta(f^*_n) \theta(f^*_d)  + \delta(x^*) \theta(-f^*_n)\theta(f^*_d),\\
        f^*_n & = 1 + g \mu M^* + \sqrt{g} \sigma \eta^*, \\
        f^*_d & = 1 - \gamma g \sigma^2 \chi^* ,\\
        f^* & = {f^*_n}/{f^*_d}.
        \label{eq:Pgeta-of-x}
    \end{aligned}
\end{equation}
We write it out in this way to make the following conditions explicit. First, $1- g \gamma \sigma^2 \chi^*$ should always be positive. {This is to be consistent with the assumption of stationarity of $x^*$, a negative value of $f_d^*$ would lead to a divergence in the dynamical equation \eqref{eq:fixed-point-approach-xt} }. This fact is non-trivial in the degree heterogeneous case, since this quantity depends on $g$ and it could possibly unbounded depending on $\nu(g)$. This immediately shows that these equations for $\gamma>0$ and $\chi^*$ can never have solutions for unbounded degree distributions. The heterogeneity alone can completely remove the possibility of having a fixed point in the asymptotic limit. The other condition is the sign of $1 + g \mu M^* + \sqrt{g}\sigma \eta^*$, which determines the survival of the species. In this case it depends also on the random variable $\eta^*$, which means that survival is not determined alone by the degree $g$, as expected due to the randomness controlled $\sigma$. Notice that it is explicitly independent from $\chi^*$ and $\gamma$.

From \eqref{eq:Pgeta-of-x} and the long time limit of \eqref{eq:M-cavity}-\eqref{eq:G-cavity} we can derive equations for $M^*$, $C^*$ and $\chi^*$,
\begin{equation}
    \begin{aligned}
        M^* & = \int \rmd g \, \nu(g) \, g \, \rmd \eta^* P(\eta^*) f^* \theta(f^*_n) \theta(f^*_d)\\
        C^* & = \int \rmd g \, \nu(g) \, g \, \rmd \eta^* P(\eta^*) {f^*}^2 \theta(f^*_n) \theta(f^*_d)\\
        \chi^* & = \int \rmd g \, \nu(g) \, g \, \rmd \eta^* P(\eta^*) \frac{1}{\sqrt{g}\sigma} \frac{\partial f^*}{\partial \eta^*}\theta(f^*_n) \theta(f^*_d)
    \end{aligned}
    \label{eq:fixed-point-MCX-equations}
\end{equation}
At this point, looking ahead to the analysis of the diverging case, we can make a change of variable that will allow us to write all interesting observables in a way that naturally continues in the diverging regime. We introduce the variables
\begin{align}
    g_c & = -\frac{1}{\mu M^*}, \hspace{10mm} q^*  = \frac{C^*}{(\mu M^*)^2}.
\end{align}

We have reused the definition of $g_c$, in this case we should think about it as a \emph{crossover} value instead of a critical one. We will show it marks the transition between nodes with higher to lower probability of surviving, meaning it plays the same role but in a soft way. With these definitions we can rewrite the main set of equations in terms of $g_c,q^*$ and $\chi^*$ as
\begin{equation}
    \begin{aligned}
        1 & = \abs{\mu} \int_0^{g^*} \rmd g \nu(g) g \frac{\sqrt{g q^*}\sigma}{1 - g \gamma \sigma^2 \chi^*}\int_{-{\Delta_g}}^{\infty} D z \, ({\Delta_g} + z),\\
        1 & = \int_0^{g^*} \rmd g \nu(g) g \frac{g\sigma^2}{(1 - g \gamma \sigma^2 \chi^*)^2}\int_{-{\Delta_g}}^{\infty} D z \, ({\Delta_g} + z)^2,\\
        \chi^* & = \int_0^{g^*} \rmd g \nu(g) g \frac{1}{1 - g \gamma \sigma^2 \chi^*}\int_{-{\Delta_g}}^{\infty} Dz, \\
        {\Delta_g} & = \sgn(\mu) \frac{g - g_c}{\sqrt{g q^*} \sigma}, \hspace{5mm} g^* = \frac{1}{\abs{\gamma} \sigma^2 \chi^* \theta(\gamma)}.
    \end{aligned}
    \label{eq:fixed-point-cavity}
\end{equation}
{where $Dz = \rme^{-\frac{1}{2}z^2}/\sqrt{2\pi} \diff z$ is the standard Gaussian measure.}

We introduce the upper limit $g^*$ simply to emphasize that equations \eqref{eq:fixed-point-cavity} might not be properly defined if $\nu(g)$ is unbounded, as $g_{\textrm{max}} < g^*$ should always be satisfied. Notice that for $\gamma=0$ we always have that $g^*\to\infty$, meaning that in this case we do not have such restriction. Regarding the role of $g_c$, we see that it sets the point where $\Delta_g$ changes sign, which is precisely controlling the survival probability, given by $\int_{-\Delta_g}^\infty Dz$. The overall sign of $\Delta_g$ is directly given by $\sgn(\mu)$, which is consistent with the intuition that in cooperation large values of $g$ are favored, and viceversa.

The final formula for the distribution of abundances at the fixed point is
\begin{equation}
    \label{eq:equilibrium-distribution}
    \begin{aligned}
        Q(x^*) & =  \int_{0}^{g^*} \rmd g \, \nu(g) \, \calN(x^*; m_g,s_g^2) \theta(x^*) + \delta(x^*) \int_{0}^{g^*} \rmd g \, \nu(g) \, \int_{-\infty}^{-{\Delta_g}} Dz ,\\
        m_g & = \frac{g_c - g }{g_c(1 - g \gamma \sigma^2 \chi^*)},\hspace{10mm}
        s_g = \frac{\sqrt{g q^*} \sigma}{\abs{g_c}(1 - g \gamma \sigma^2 \chi^*)}.
    \end{aligned}
\end{equation}
{where $\calN(x^*;m,s^2)$ denotes a Gaussian distribution for $x^*$ with mean $m$ and variance $s^2$}.
We can see that this result is a generalization of \eqref{eq:equilibrium-minimal-model}, where we now have a mixture of truncated Gaussian distributions. The effect of the network heterogeneity is such that the means $m_g$ increase with degree if $\mu>0$ and the opposite for $\mu<0$. The variance $s_g^2$ always increases with $g$.

To explore the diverging regime we need to find the critical values where $M^*$ diverges, which can be found from \eqref{eq:fixed-point-cavity} by setting $g_c = 0$. For $\mu>\mu_c$ we can assume $y(\tau)$ reaches a fixed point and that the effective dynamics also simplifies close to the fixed point,
\begin{equation}
    \dot y \approx y(\tau)(-y(\tau) + g  + \sqrt{g}\sigma \xi^* + g \gamma \sigma^2 \chi^* y(\tau) - \mu B^*),
\end{equation}
where $\xi^*$ is a Gaussian zero mean random variable with variance $\expected{{\xi^*}^2} = \expected{{y^*}^2} = q^*$ $\chi^*$ is the same as in \eqref{eq:integrated-response}.
Following similar arguments as before, we can write a set of equations for $B^*$, $q^*$ and $\chi^*$. We observe that again reusing the definition for $g_c = \mu B^*$, the equations are the same as in the the previous regime. That is, in terms of $g_c$, $q^*$, and $\chi^*$ the set of equations \eqref{eq:fixed-point-cavity} are the same for all values of $\mu$, whether the system diverges or not. The only exception is $\mu =0 $, where $M^*$ does not play a role, but the equations can be easily derived from \eqref{eq:fixed-point-MCX-equations}.

Once these equations are solved, we can calculate the final distribution to be given by
\begin{equation}
\begin{aligned}
    Q(y^*) & = \int_{0}^{g^*} \diff g \nu(g) \calN(y^*; m_g,s_g^2) \theta(y^*) + \delta(y^*) \int_0^{g^*} \diff g \nu(g) \int_{-\infty}^{-{\Delta_g}} Dz,\\
    m_g & = \frac{g - g_c}{1 - g \gamma \sigma^2 \chi^*}, \hspace{10mm}s_g  = \frac{\sqrt{g q^*}\sigma }{1 - g \gamma \sigma^2 \chi^*}.
\end{aligned}
\label{eq:y-equilibrium distribution}
\end{equation}
Except for a factor of $g_c$, it is basically the same mixture of Gaussians as for the converging case. 

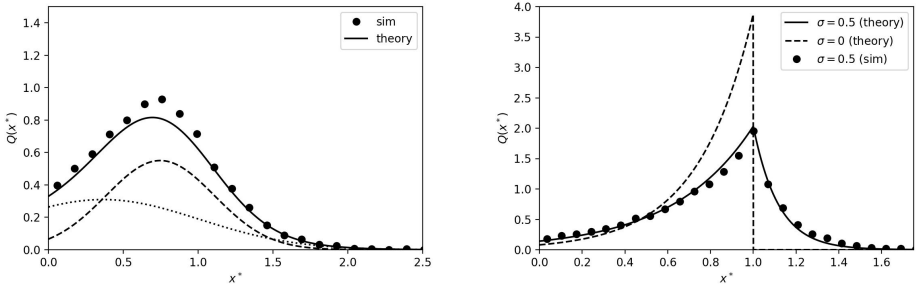
\begin{figure*}
    \centering
    \begin{picture}(370,125)
    % \put(0,0){\line(1,0){370}}
    % \put(0,0){\line(0,1){125}}
    \put(15,10){\includegraphics[trim = 30 20 0 0 ,clip,width=.49\textwidth]{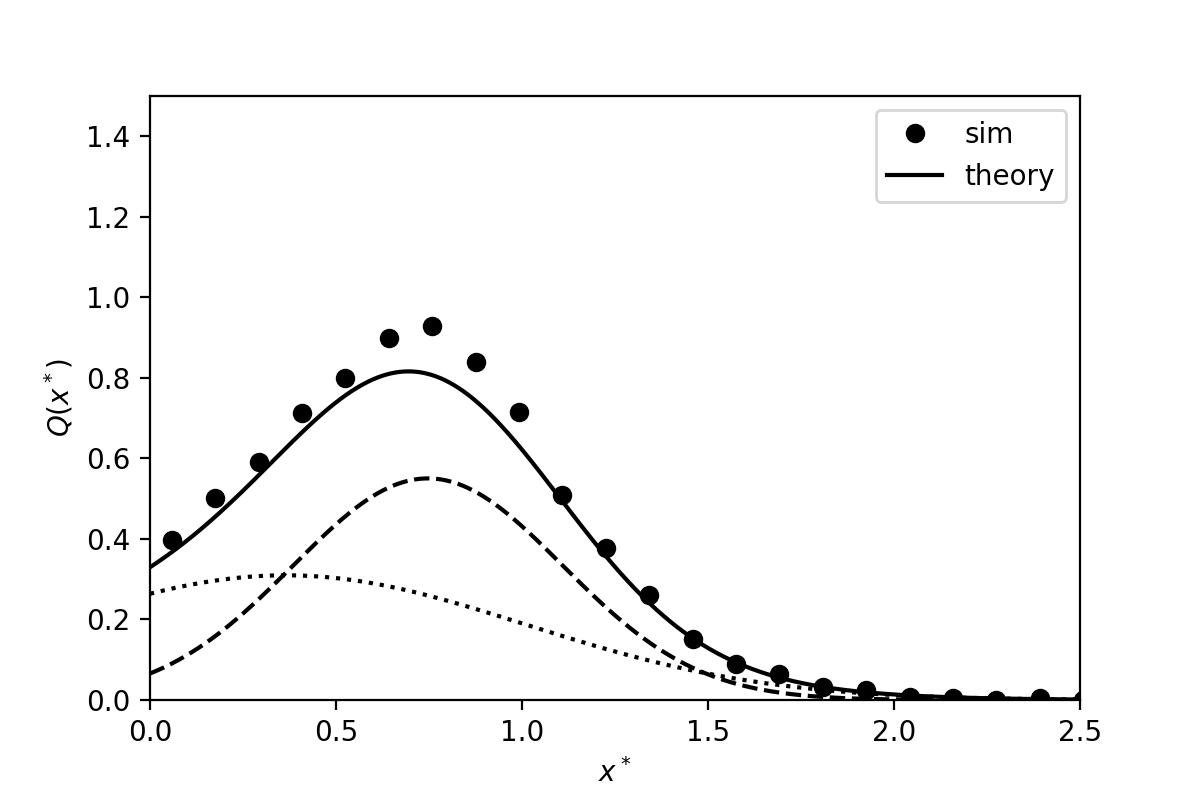}}
    \put(205,10){\includegraphics[trim = 26 15 0 0 ,clip,width=.49\textwidth]{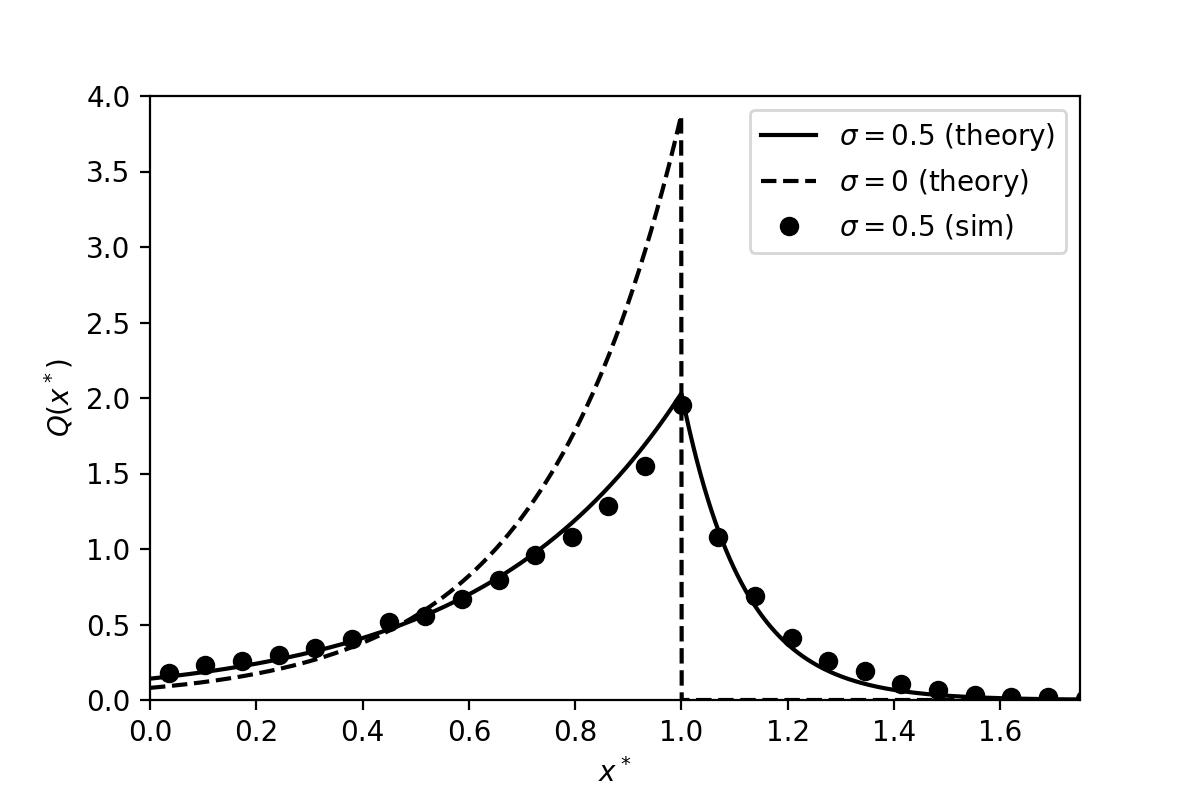}}
    \put(95,0){\scalebox{1.2}{$x^*$}}
    \put(290,0){\scalebox{1.2}{$x^*$}}
    \put(0,60){\rotatebox{90}{\scalebox{1}{$Q(x^*)$}}}
    \put(190,60){\rotatebox{90}{\scalebox{1}{$Q(x^*)$}}}
    \end{picture}
    \caption{Left: Abundance distribution for the bimodal case, \eqref{eq:bimodal-distribution}, with $N=1000$, $C = 50$, $\mu=1$, $\sigma=0.5$, and $\gamma =0.9$. Results are shown for averaging over 3000 instances. The dotted and the dashed line correspond to the theoretical distributions of each of the two degrees. We can see the full distribution corresponds to the mixture of them, solid line \eqref{eq:equilibrium-distribution}. Right: Equilibirum distribution for an exponential degree distribution. Average over 300 instances. {Error bars are of the order of magnitude of the symbols and omitted for clarity.}}
    \label{fig:fixed-points-general}
\end{figure*}

We test these results against numerical simulations and find a good agreement when exploring the bimodal ensemble, \eqref{eq:bimodal-distribution}. In Figure \ref{fig:fixed-points-general} we show comparison with numerical simulations for the abundance distribution at the fixed point \eqref{eq:equilibrium-distribution}.

In order to gain a deeper intuition of the solutions of \eqref{eq:fixed-point-cavity} for the case of $\gamma = 0$ we can make the following approximation. Assuming small $\sigma$, for $g_c$ and $q^*$ we use the same values as for the homogeneous case, $\sigma = 0$. This means using \eqref{eq:g_c-equation} for $g_c$, and calculating $q^*$ from \eqref{eq:Pg-homogeneous} we find
\begin{equation}
    q^* \approx \expected{g^3}_{g_c} - 2 \expected{g^2}_{g_c}g_c  + \expected{g}_{g_c}g_c^2,
    \label{eq:qstar-approx}
\end{equation}
We can then use this values directly in our expressions for $Q(x^*)$, 
 \eqref{eq:equilibrium-distribution}, and  $Q(y^*)$, \eqref{eq:y-equilibrium distribution}, directly. Therefore \eqref{eq:g_c-equation} and \eqref{eq:qstar-approx} constitute the small $\sigma$ approximation for \eqref{eq:fixed-point-cavity} in the $\gamma = 0$ case.

We test this approximation for an exponential degree distribution, $\nu(g) = \rme^{-g}$, where we can actually perform the integrals over the degree distribution exactly, giving
\begin{align}
    Q(x^*) = \left\lbrace\begin{array}{cc}
        \displaystyle\frac{g_c\exp\p{\frac{g_c}{q^* \sigma^2}( 1-x^* - \abs{1- x^*}\sqrt{1 + 2 q^* \sigma^2})}}{\sqrt{1 + 2 q^* \sigma^2}} & \textrm{if } \mu<0 \\
    \displaystyle\frac{\abs{g_c}\exp\p{-\frac{\abs{g_c}}{q^* \sigma^2}  (1-x^* + \abs{x^*-1}\sqrt{1 + 2 q^* \sigma^2})}}{\sqrt{1 + 2 q \sigma^2 }}     & \textrm{if }0<\mu<\mu_c
    \end{array}\right.,
\end{align}
and for the diverging case,
\begin{align}
    Q(y^*) = \frac{\exp\p{-\frac{1}{q^* \sigma^2} (g_c + y^*) (-1 + \sqrt{1 + 2 q^* \sigma^2})}}{\sqrt{1 + 2 q^* \sigma^2 }}    \hspace{5mm} \textrm{if }\mu>\mu_c .
\end{align}
In Figure \ref{fig:fixed-points-general} we show the approximation works quite well even for $\sigma = 0.5$. Therefore, we see that the results even with $\sigma>0$ are very similar to the homogeneous case, $\sigma = 0$. In the competitive case, $\mu<0$, the distribution develops a cusp close to $x^*=1$ where most of the species with the least amount of competitors lie. Most of the mass is between $0$ and $1$, but there is a quickly decaying tail due to species that by luck managed to do better. In the cooperative case $\mu>0$ the distribution also has a cusp at $1$, but now most of the mass is on values with $x^*>1$. In the diverging case it is a simple exponential distribution. We see that the main difference with the homogeneous case, \eqref{eq:equilibrium-minimal-model}, is the appearance of tails below or above $g_c$ depending if one is in the cooperative or competitive regime respectively. The size of these tails is controlled by $\sigma$ and $q^*$ that can be expressed in terms of statistics of the degree distribution. Importantly, we wish to point out that the intuition and the role of $g_c$ is maintained in this regime.

\begin{figure*}
    \centering
    \begin{picture}(370,125)
    % \put(0,0){\line(1,0){370}}
    % \put(0,0){\line(0,1){125}}
    \put(15,10){\includegraphics[trim = 30 20 0 0 ,clip,width=.49\textwidth]{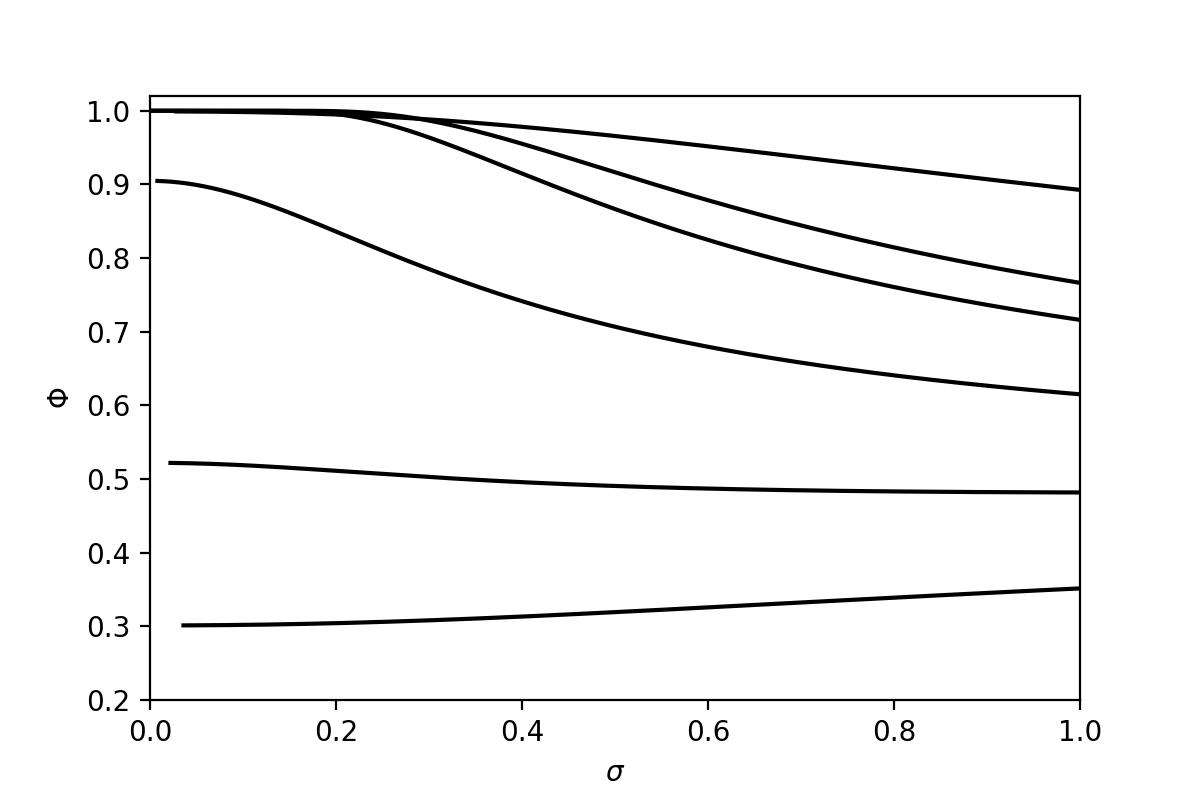}}
    \put(200,10){\includegraphics[trim = 26 15 0 0 ,clip,width=.49\textwidth]{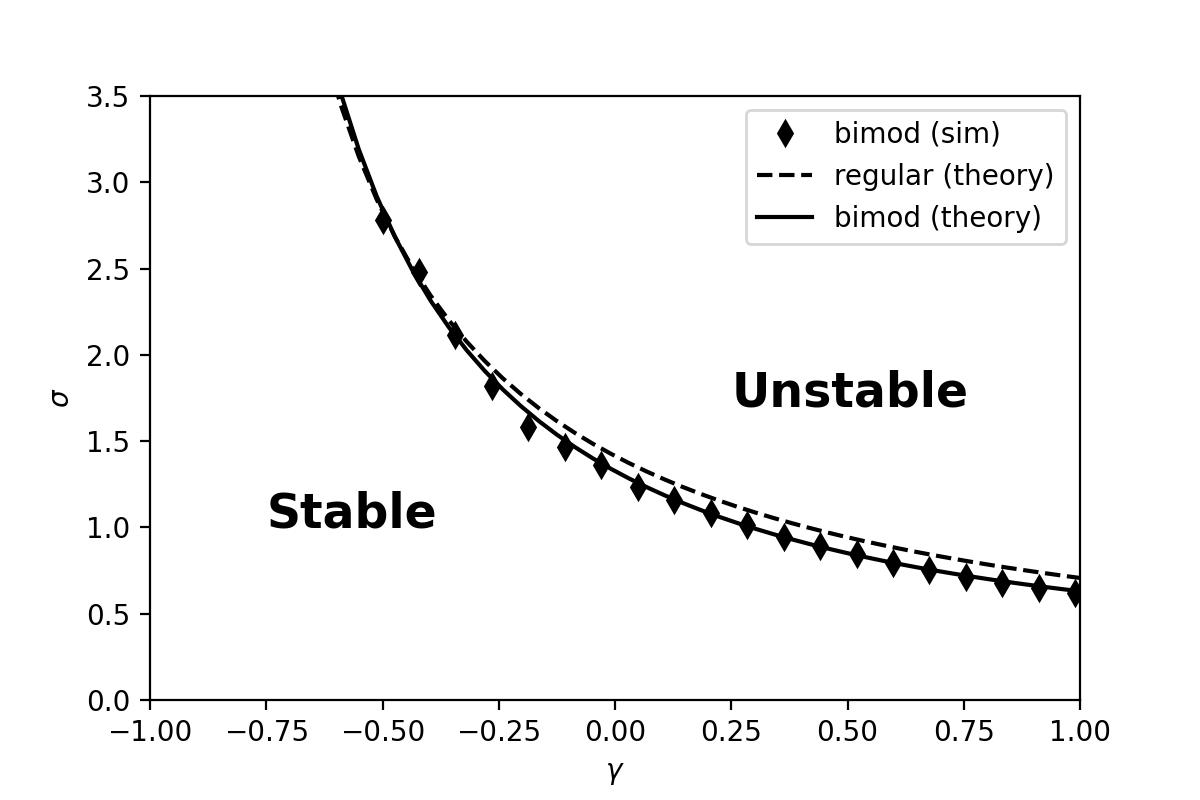}}
    \put(100,0){\scalebox{1.2}{$\sigma$}}
    \put(283,0){\scalebox{1.2}{$\gamma$}}
    \put(0,60){\scalebox{1.2}{$\Phi$}}
    \put(190,60){\scalebox{1.2}{$\sigma$}}
    \end{picture}
    \caption{Left: Survival probability, $\Phi$, for $\gamma=0$ and exponential distribution as a function of $\sigma$, from \eqref{eq:survival-probability}. From top to bottom different values $\mu=$ $-.2$, $0.33$, $0.4$, $0.5$, $0.7$, and $1$. Notice for $\mu = 0.7$ it is non-monotonic and for $\mu = 1$ it is an increasing function. Right: Phase boundary for the linear stability described by the HDMFT. Dashed line corresponds to the theoretical values, from \eqref{eq:critical-line-sigma}, for non-heterogeneoud DMFT, $\textrm{VAR}[g] = 0$, and solid line for the bimodal degree distribution, \eqref{eq:bimodal-distribution}. The phase boundary made by the solid diamonds was determined numerically from \eqref{eq:Jacobian} for $N=2000,C = 66$. {Error bars are of the order of magnitude of the symbols and omitted for clarity.}}
    \label{fig:survival_phase_diagram}
\end{figure*}
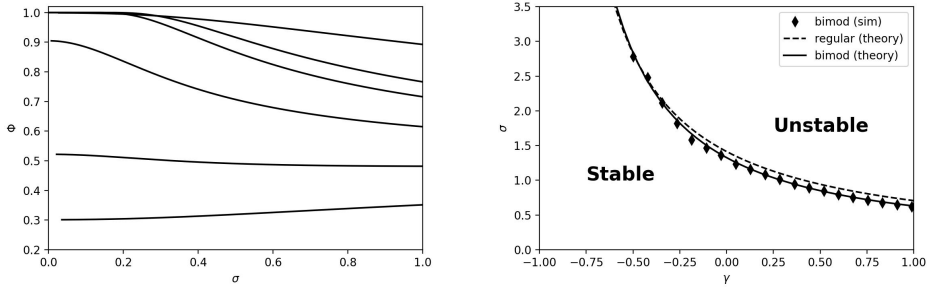

\subsection{Survival probability and Gini coefficient}

We can write down the probability of survival as a function of the degree. This can be calculated in both regimes with the same formula, 
\begin{equation}
\begin{aligned}
    \phi(g) & = \int_{-{\Delta_g}}^\infty Dz \\
    {\Delta_g} & = \frac{g - g_c}{\sqrt{g q^*} \sigma} \sgn(\mu)
\end{aligned}
    \label{eq:survival-probability-by-degree}
\end{equation}
where we only need to remember that the interpretation of extinction is not the same for all values of $\mu$. When $\mu>0$ and $g_c>0$, this fraction $\phi(g)$ refers to the relative survival, as all quantities diverge in this regime. We can see that the role of the heterogeneity in strengths, $\sigma$, is that of smoothing the survival probability, tending to a step function as one approaches the homogeneous case when $\sigma \to 0$.

The overall survival probability is given by the integral over the degree distribution,
\begin{equation}
    \Phi = \int_0^{g^*} \diff g \, \nu(g)\, \phi(g)
\end{equation}

To gain more insight of the effect of the degree heterogeneity, we present the solution for $\gamma = 0$ and exponential degree distribution $\nu(g) = \rme^{-g}$. In this case the integral can be performed exactly, 
\begin{equation}
\begin{aligned}
    \Phi &= \left\lbrace \begin{array}{cc}
        1 + \frac{1}{2} \p{-1 -\frac{1}{f}} \exp\p{\frac{g_c}{q^* \sigma^2}(1-f)} &  \textrm{if } \mu<0 \\
        1 + \frac{1}{2} \p{-1 +\frac{1}{f}} \exp\p{\frac{g_c}{q^* \sigma^2}(1+f)} &\textrm{if } 0<\mu<\mu_c   \\
        \frac{1}{2} \p{1 + \frac{1}{f}} \exp\p{\frac{g_c}{q^* \sigma^2}(1-f)} & \textrm{if } \mu_c<\mu 
    \end{array}\right.\\
    f & = \sqrt{1 + 2 q^* \sigma^2}
\end{aligned}
\label{eq:survival-probability}
\end{equation}
Using the small $\sigma$ approximation we can plot the survival probability for different value of $\mu$ in Figure \ref{fig:survival_phase_diagram}.

We can therefore also calculate the Gini coefficient in this case. We find that for $\mu<\mu_c$ the Gini coefficient \emph{increases} with $\sigma$, that is the heterogeneity makes the distribution more unequal. Nevertheless, in the diverging case, $\mu>\mu_c$, we find there is a change in behavior. The formula under the small $\sigma$ approximation for the Gini coefficient in this case is
\begin{equation}
\begin{aligned}
    G(\mu,\sigma) & = 1 - \frac{1}{4}\p{1 + \frac{1}{w(\mu,\sigma)}}\exp\p{\displaystyle\frac{2(2 + W_{-1}(-\rme^{-2}/\mu)}{1 + w(\mu,\sigma)}}\\
    w(\mu,\sigma) & = \sqrt{1 + \frac{4 \sigma^2}{\mu} \p{1-\frac{1}{W_{-1}(-\rme^{-2}/\mu)}}}.
\end{aligned}
\end{equation}
{where $W_{-1}(x)$ corresponds to the Lambert W function.}

Interestingly it can display non-monotonous behaviour with respect of $\sigma$, if we define the equation 
\begin{equation}
    \frac{\partial G}{\partial \sigma} {(\mu^\star(\sigma),\sigma)}= 0,
\end{equation}
we find there is a critical line coming out of the $\sigma = 0$ axis at $\mu^\star(0) = \rme/3$, which coincides precisely with $g_c = 1$. More importantly, for all parameter combinations $(\mu,\sigma)$ on the right of the $(\mu^\star(\sigma),\sigma)$ line the Gini coefficient decreases with heterogeneity instead of increasing. 

This might seem a bit as a counterintuitive result, how is it that increasing the heterogeneity reduces the inequality of the longtime distribution? The reason for it comes from the fact that in this high $\mu$ regime fluctuations can save you from extinction. So for $\mu>\rme/3$, even though heterogeneity is making some species more abundant, it is saving more from relative extinction and therefore the inequality reduces. This can be seen from the fact that we obtain the same phase diagram if we look at the derivative of the survival probability, which has exactly the same boundary for equation
\begin{equation}
    \frac{\partial \Phi}{\partial \sigma} = 0
\end{equation}
The change in behavior can be observed in Figure \ref{fig:survival_phase_diagram}, where we see that for low values of $\mu$ the survival probability is decreasing with $\sigma$ but increasing for higher values.

\section{Linear stability analysis \label{sec:linear}}

To study the stability conditions of the fixed point described by \eqref{eq:equilibrium-distribution} we linearize the original model around a given equilibrium fixed point $x_i = x^*_i + \epsilon_i$, the dynamics of the linearized model are given by, 
\begin{equation}
    \dot{\epsilon}_i = x^*_i ( -\epsilon_i + \sum_j A_{ij} \alpha_{ij} \epsilon_j).
\end{equation}
This means that the stability is determined by the eigenvalues of a random matrix of the following structure
\begin{equation}
    J_{ij} = x_i^* (-\delta_{ij} + A_{ij}\alpha_{ij}),
    \label{eq:Jacobian}
\end{equation}
The system will become unstable when $J_{ij}$ develops eigenvalues with positive real part.

If we consider a degree distribution $\nu(g)$ with small degree heterogeneity, that is $\textrm{Var}[g]<1$, one can make an expansion in the variance and obtain the edges of the eigenvalue distribution of $-\delta_{ij}+ A_{ij}\alpha_{ij}$. They were calculated this way in \cite{baron2022eigenvalue}, giving the formula
\begin{equation}
    \begin{aligned}
    \lambda_{\textrm{max}} & = -1 + \sigma ( 1 + \gamma) (1 + \frac{\delta^2}{2} (1 + 2 \gamma - \gamma^2))\label{eq:edge-formula}\\
    \delta^2 & = \textrm{Var} [g]
    \end{aligned}
\end{equation}
Nevertheless, this is not taking into account the fact we are multiplying by $x_i^*$, which has some impact on the eigenvalues, especially because a large fraction of this numbers might be zero due to extinction. Nevertheless, one can get away with a simple calculation if one is interested only on the edge \cite{galla2018dynamically}. It is only necessary to scale the variance according to the fraction of surviving species and use this value in \eqref{eq:edge-formula}
\begin{equation}
    \sigma^2 \longrightarrow \Phi \, \sigma^2.
\end{equation}
This leads to the critical line by enforcing the condition $\lambda_{\textrm{min}} = 0$, 
\begin{equation}
    \sigma_c (\gamma) = \sqrt{\frac{1}{\Phi}} \frac{1}{( 1 + \gamma) (1 + \frac{\delta^2}{2} (1 + 2 \gamma - \gamma^2))}
    \label{eq:critical-line-sigma}
\end{equation}
We show in Figure \ref{fig:survival_phase_diagram} a good match of this formula with numerical simulations.  

{This result can also be derived starting from the HMDFT \eqref{eq:effective-process} and performing a linear stability analysis around the fixed point given by \eqref{eq:fixed-point-cavity}, a procedure originally described in \cite{opper1992phase}. Such analysis is performed in \cite{park2024incorporating}}. An extended analysis of the linear instability is beyond the scope of this paper.

\section{Beyond mean-field \label{sec:beyond-mean-field}}

So far we have only described the mean-field results, obtained by assuming $C\gg1$. It is natural to ask what are the non mean-field effects and also \emph{when} do we expect to observe them. It is natural to expect them at small values of $C$, since fluctuations should not be discarded. Obviously, $C$ should be small compared to $N$, and this is typically the case. What is less obvious is how $C$ should compare to the other parameters of the model.

\begin{figure*}
    \centering
    \begin{picture}(370,125)
    % \put(0,0){\line(1,0){370}}
    % \put(0,0){\line(0,1){125}}
    \put(15,10){\includegraphics[trim = 30 20 0 0 ,clip,width=.49\textwidth]{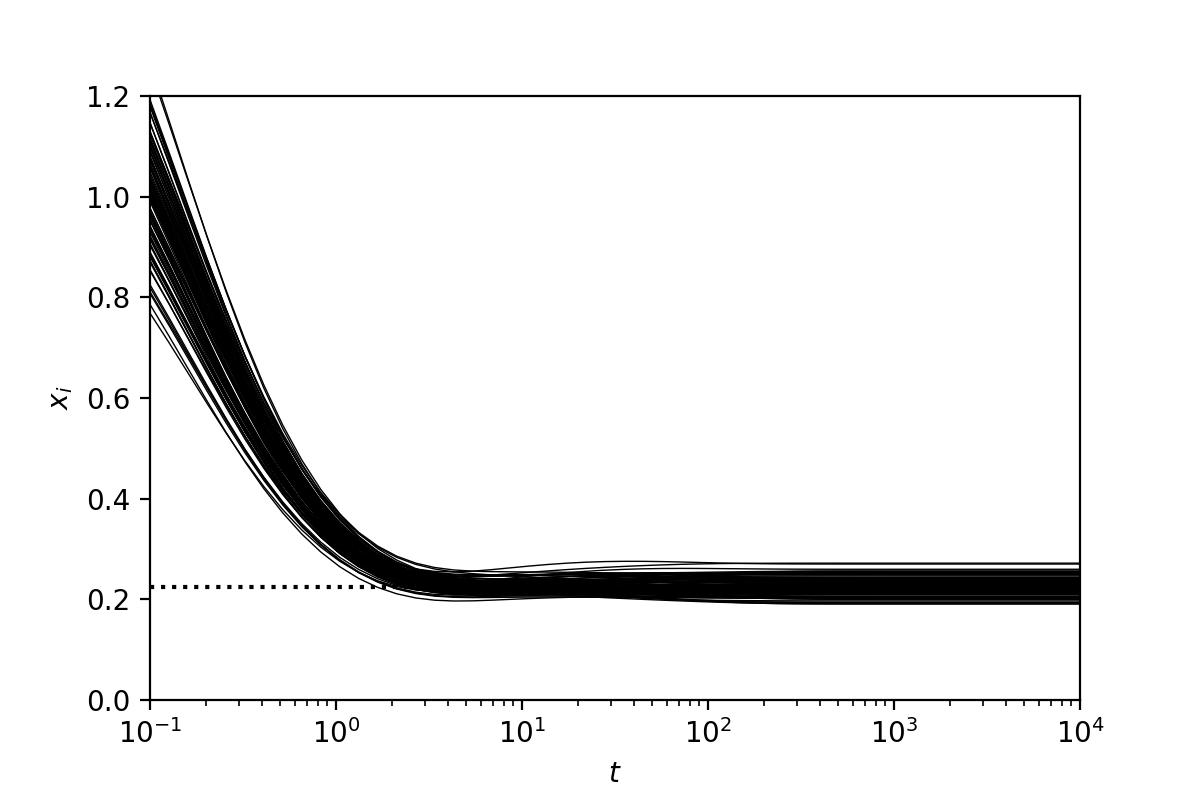}}
    \put(200,10){\includegraphics[trim = 26 15 0 0 ,clip,width=.49\textwidth]{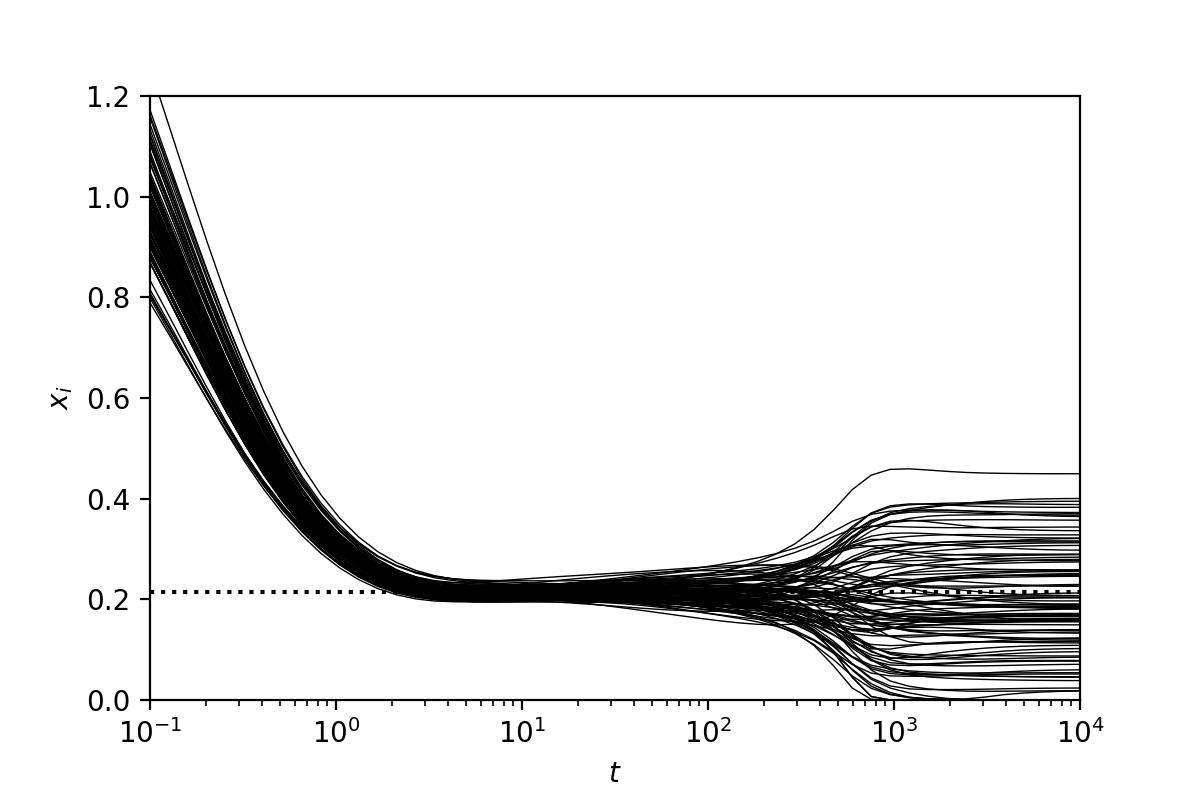}}
    \put(100,0){\scalebox{1.2}{$t$}}
    \put(283,0){\scalebox{1.2}{$t$}}
    \put(0,70){\scalebox{1.2}{$x_i$}}
    \put(185,70){\scalebox{1.2}{$x_i$}}
    \end{picture}
    \caption{Both plots show solutions of the homogeneous model \eqref{eq:model-competition} on a regular graph with $N = 30000$ and $C=50$. For this graph the stability threshold value is $\mu_{st} \approx -3.57$. Only a sample of $100$ species is shown. Dotted line indicates the mean-field fixed point, $1/(1-\mu)$. Left: $\mu = -3.56$, above the stability threshold, therefore it follows closely the mean-field. Right: $\mu = -3.58$, just below the stability threshold, there it diverges from the mean-field solution after long time.  }
    \label{fig:beyond-mean-field}
\end{figure*}
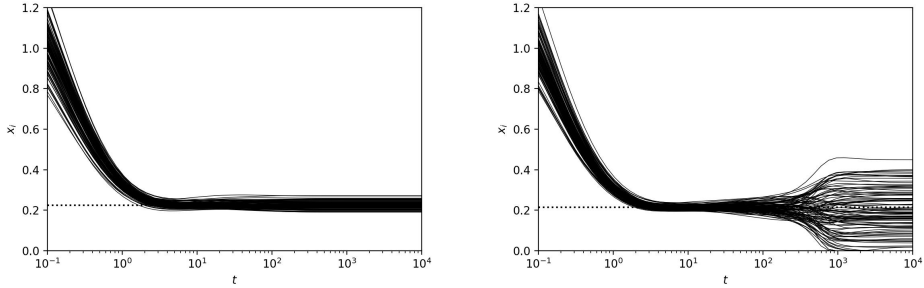

Without going into full detail, we will simply give a flavor of when and how deviations occur. The theory will always fail to account for fluctuations of order $1/\sqrt{C}$, but this is not so problematic as the means can still be well described by the theory, as shown in Figure \ref{fig:dynamics}. The big differences occur when the fixed point described at the mean-field level is no longer stable in the finite $C$ case. As an example we can take the easiest possible instance, a random regular graph with no degree heterogeneity of any kind, $\nu(g) = \delta(g-1)$ and $\sigma = 0$. In this case it is easy to prove that the mean-field fixed point will be unstable in the highly competitive regime if $\mu<-C/2/\sqrt{C-1}$, as it is done in \cite{marcus2022local} (presented here in our notation) by looking at the exact edges of the eigenvalue distribution of the Jacobian. This gives us an indication that we can define $\mu_{st} = -C/2/\sqrt{C-1}$ for $\mu>\mu_{st}$ the DMFT should describe average quantities correctly. If we now consider a stronger interaction strength, $\mu<\mu_{st}$ we do not expect the theory to describe the system. Nevertheless, if $\mu$ is more negative but still close to $\mu_{st}$, then we can see that the DMFT can describe the transient regime. This is due to the fact that the unstable directions have positive but small eigenvalues associated with them, making the instability taking very long to develop. This can be seen in Figure \ref{fig:beyond-mean-field}.

In the general case, for an arbitrary degree distribution the edges of the eigenvalue distribution might scale differently with $C$, \cite{kuhn2008spectra,rogers2008cavity}, but overall we can conclude that this will determine the range of $\mu$ on which the mean-field will be a reasonable approximation or not. 

Regarding the inequality of the distributions, there is a big difference between the mean-field case and the finite $C$. In the mean-field case inequality is strongly dependent on the inequality of the degree distribution $\nu(g)$. In particular in the homogeneous case, $\sigma=0$, if the degree distribution is non-heterogeneous, $\nu(g) = \delta(g-1)$, there is no inequality at all. This is certainly not the case for values of $\mu$ beyond the mean-field limit, as proved in \cite{marcus2022local} and show in Figure \ref{fig:beyond-mean-field}. There can be many extinctions even if there is no degree heterogeneity. A picture totally absent in the mean-field case. We conjecture something similar should happen in the highly cooperative regime.

\section{Discussion \label{sec:discussion}}

In this paper we have shown how to deal with the generlizaed Lotka-Volterra model for a high number of species interacting in a heterogeneous way through a random network. Focusing in the asymptotic case where not only the number of species is large, $N\gg1$, but also the $C\gg 1$ case. We have shown that at long times the system relaxes to a fixed point of which we can provide an statistical description with the help of HDMFT. We have developed a general theory for a large class of degree distributions, $\nu(g)$, and for arbitrary values of $\mu$, $\sigma$, and $\gamma$ (provided the mean-field fixed point is stable). While the effective process \eqref{eq:effective-process} should provide a general description of the model, it is hard to extract analytically information from it. When looking only at the fixed point, \eqref{eq:fixed-point-cavity}, the theory is much more manageable and interpretable. Even though the equations do not have a simple explicit analytic solution, the formulae for the asymptotic distributions, \eqref{eq:equilibrium-distribution} and \eqref{eq:y-equilibrium distribution}, allow to extract a lot of information about the final state of the system. Furthermore, we have shown that in the case of no correlations between interaction weights, $\gamma = 0$, the general theory is well approximated as a perturbation of the homogeneous case. We know that in this case, the main object is the critical degree, $g_c$, above or below which go extinct depending if interactions are mostly competitive or mostly cooperative respectively. Once weight heterogeneity is incorporated by taking $\sigma>0$, the effect is only to soften the theory, turning $g_c$ into a crossover value. The simple expressions \eqref{eq:g_c-equation} and \eqref{eq:qstar-approx} for order parameters $g_c$ and $q^*$ can be used for all values of $\mu$ and a wide range of values of $\sigma$ even in the exponential case that already has non-trivial degree heterogeneity. With this approximation we have been able to show an interesting change in the behavior of the survival probability with $\sigma$ for the exponential case.

Nevertheless, we have only considered the case of a unique stable fixed point. That is, for values below the critical line \eqref{eq:critical-line-sigma}. It is well known there is a much richer behaviour above it, see \cite{bunin2017ecological,altieri2021properties,de2024many,roy2019numerical} for examples. Understanding how the heterogeneity of the network structure changes such a regime is a very much open problem, even in the heterogeneous mean-field case. Even for other systems like spin glasses, \cite{metz2022mean}, only the replica symmetric regime has been explored. 

As mentioned in the last section, understanding the difference with the finite $C$ case is a non-trivial task. One basic question is whether finite connectivity effects increase or decrease the inequality. In Figure \ref{fig:beyond-mean-field} we can clearly see inequality increases in the finite $C$ regime, but is there a case where it is the opposite? In general we conjecture that in the mean-field case the ranking is completely correlated with the degree $g$, \eqref{eq:survival-probability}. Beyond the mean-field this should not be the case, as there might be dependence on initial conditions or on other non-local properties of the nodes. While precise knowledge of initial conditions might be generally impossible or irrelevant for theoretical ecology, this is of great interest for the study of ABM's. Knowing the effect of initialization of large ABM's is of crucial importance for their large scale application, \cite{kolic2022estimating}. Overall we conjecture that understanding the difference between \emph{mean-field} behavior and whatever lies \emph{beyond mean-field} one is of paramount importance for the correct utilization and interpretation of ABM's. See \cite{patil2023stability} for an interesting example.

Further directions of improvement are considering correlations with additional features present in the nodes,\cite{annibale2009tailored}. Also, applying other techniques to follow the dynamics at the finte $C$ case, \cite{mozeika2008dynamical,mozeika2009dynamical} to the Lotka-Volterra should be possible. Additionally, other random graph ensembles that take into account the presence of short loops in the network could be considered, \cite{lopez2018exactly,lopez2020imaginary,lopez2021transitions,foster2010communities,newman2009random}.

The author acknowledges that two weeks before and just a few days before, the preprints \cite{park2024incorporating} and \cite{poley2024interaction} appeared respectively in \texttt{arxiv.org}. They both have similar but not completely overlapping results between themselves and with this work.

\section*{Acknowledgements}
The author thanks Jean-Phillipe Bouchaud, Jerome Garnier-Brun and Michael Benzaquen for careful discussions during the development of this project, as well as Matteo Smerlak, Ada Altieri, James P. Sethna, Valentina Ros, Mauro Pastore, and Silvio Franz for discussions and feedback.

This research was conducted within the Econophysics \& Complex Systems Research Chair, under the aegis of the Fondation du Risque, the Fondation de l’\'Ecole polytechnique, the \'Ecole polytechnique and Capital Fund Management.

\section*{References}
\bibliography{refs}

\appendix

\section{Generating functional analysis \label{app:generating-functional}}

We focus on studying the distribution of the dynamical trajectory followed by a randomly chosen species. For a given instance we denote this distribution by $Q_N[x]$. In general we will use square brackets to denote a functional dependence, that is 
\begin{equation}
    Q_N[x] = \frac{1}{N} \sum_{i=1}^N \prod_{t>0} \delta(x(t) - x_i(t; \bx_0, \bA, \balpha )) = \frac{1}{N}\sum_{i=1}^N \delta[x - x_i(\bx_0, \bA, \balpha)],
    \label{eq:empirical-path-distribution}
\end{equation}
where $x_i(t; \bx_0, \bA, \balpha )$ denotes a solution of \eqref{eq:model-competition} for a given initial condition $\bx_0$, given interaction matrix $\bA$, and given interaction strengths $\balpha$.

As it is typical in statistical physics, the objective is to calculate the average distribution in the thermodynamic limit, that is
\begin{equation}
    Q[x] = \lim_{N\to \infty} \overline{Q_N[x]}
\end{equation}
where $\overline{ \cdot}$ denotes an average over $\bx_0$, $\bA$, and $\balpha$. In order to calculate the statistical properties of this distribution, it is typical to work with the dynamical moment generating functional, also known as the Martin-Siggia-Rose-Janssen-de Dominicis path integral (MSRJD), \cite{martin1973statistical,de1978dynamics}. It is defined by 
\begin{equation}
    Z[\bpsi] = \overline{ \int \calD \bx \, \delta[\textrm{equations of motion}] \, \rme^{ \rmi \bx \cdot \bpsi}},
    \label{eq:dyn-generating-function}
\end{equation}
where we define the dot product between vector paths to be defined as
\begin{equation}
    \begin{aligned}
        x \cdot \psi = {} & \int \diff t \, x(t) \psi(t)\\
        \bx \cdot \bpsi = {} & \sum_{i=1}^N x_i \cdot \psi_i
    \end{aligned}
\end{equation} 
We will also use the symbol $\calD$ for a functional measure in the space of paths.
\begin{equation}
    \calD \bx = \prod_{i=1}^N \calD x_i = \prod_{i=1}^N \prod_{t>0} \diff x_i(t)
\end{equation}

While there exists a lot of literature to deal with objects such as \eqref{eq:dyn-generating-function} in the mean-field case, it is generally impossible to do so exactly in the case when the graph has a finite average degree, $C = \Order{1}$. To overcome this problem withouth going back to the typical mean-fiel case, we look at the so called heterogeneous mean-field limit. Analytically this method consists of taking two different asymptoticl limtis, $N\to \infty$ and \emph{afterwards} $C\to \infty$. This is equivalent to consider systems where $C$ is growing sublinearly with $N$, for example $C = \Order{\log N}$. 

We can work with the generating function by using the Fourier representation of the functional Dirac delta,
\begin{equation}
    \begin{aligned}
        Z[\bpsi] ={}& \int \calD x \, \calD \hx \, \exp\p{\rmi \bx \cdot \bpsi} \\
        & \times \exp\p{ \rmi \sum_i \int \diff t \, \hx_i(t) \p{\frac{\dot{x}_i(t)}{x_i(t)} - \s{1 - x_i(t) + \sum_j A_{ij} \alpha_{ij} x_j(t) + h(t)}  }}
    \end{aligned}
\end{equation}

We know look to average over the disorder, $\overline{Z[\bpsi]}$, where the disorder corresponds to both the edge the disorder and the whole graph $\bA$. We separate the only term that depends on the disorder,
\begin{equation}
    \begin{aligned}
        \overline{Z[\bpsi]} = {} &\int \calD \calM \, \overline{\exp\p{- \rmi \sum_{ij} \int \diff t \, A_{ij}   \alpha_{ij} x_j(t) \hx_i(t)  - \rmi \sum_{ij} \omega_i A_{ij}}} \\
        \calD\calM = {} &  \frac{1}{\calZ}  \calD \bx \calD \hx \diff \bomega \, \exp\p{\rmi \bpsi \cdot \bx + \rmi \hbx \cdot\p{ \frac{\dot{\bx}}{\bx} - \s{ 1 - \bx + \boldsymbol{h}}} + \rmi \bomega \cdot \boldsymbol{k}}
    \end{aligned}
\end{equation}
We have used the standard trick to transform an average over configuration model to an average over the ER ensemble that has independent edges. The new fields come from the Fourier representation of the Kroenecker deltas, $\delta_{nm} = \int_0^{2\pi} \diff \omega\, \rme^{\rmi \omega (n-m)}$. We can exploit this fact to proceed analytically. 
\begin{equation}
    \begin{aligned}
        &\overline{ \exp\p{- \rmi \sum_{ij} \int \diff t \, A_{ij}   \alpha_{ij} x_j(t) \hx_i(t)  - \rmi \sum_{ij} \omega_i A_{ij}}} = \\
        & = \overline{\exp\p{-\rmi \sum_{i<j} A_{ij} \s{\omega_i + \alpha_{ij} x_j \cdot \hx_i  + \alpha_{ji} x_i \cdot \hx_j + \omega_j}}} \\
        & = \prod_{i<j} \overline{\exp\p{-\rmi  A_{ij} \s{\omega_i + \alpha_{ij} x_j \cdot \hx_i  + \alpha_{ji} x_i \cdot \hx_j + \omega_j}}}\\
        & = \prod_{i<j} \p{1 + \frac{C}{N} \overline{\rme^{- \rmi \s{\omega_i + \alpha_{ij} x_j \cdot \hx_i  + \alpha_{ji} x_i \cdot \hx_j + \omega_j} } - 1}}\\
        & = \exp\p{- \frac{NC}{2} + \frac{C}{2 N} \sum_{ij}  \rme^{- \rmi \omega_i}\overline{\exp\p{- \rmi \s{\alpha x_j \cdot \hx_i  + \alpha' x_i \cdot \hx_j } }} \rme^{- \rmi \omega_j }},
    \end{aligned}
\end{equation}
where the overline $\overline{\cdot}$ is only an average over $(\alpha,\alpha')$. We now introduce the next functional order parameter.
\begin{equation}
    P[x,\hx] = \frac{1}{N}\sum_{i=1}^N \delta\s{x - x_i} \delta\s{\hx - \hx_i}\rme^{-\rmi \omega_i}
\end{equation}

Disregarding constants, we get can use functional Dirac deltas to write the generating function in the following way, 
\begin{equation}
    \begin{aligned}
        \overline{Z[\bpsi]} \propto {} & \int \calD P \calD \hP \, \rme^{N S[P,\hP]} \sim \rme^{N S[P^\star, \hP^\star]}\\
        S[P,\hP] = {} & \rmi P \cdot \hP + \frac{C}{2} P\cdot \bW P + \sum_k p(k) \log\frac{C^k \calA(k)}{k!} \\
        P \cdot \hP = {} & \int \calD x \calD \hx P[x,\hx] \hP[x,\hx] \\
        P\cdot \bW P = {} & \int \calD x \calD \hx  \calD y \calD \hy P[x,\hx] P(y,\hy) \overline{\exp\p{- \rmi \s{\alpha y \cdot \hx  + \alpha' x \cdot \hy }}} \\
        \calA(k) = {} & \int \calD x \calD \hx \exp\p{\rmi x \cdot \psi_k  + \rmi \hx \cdot \p{\frac{\dot x}{x} - [1 - x + h]}} \p{-\frac{i \hP(x, \hx)}{C}}^k
    \end{aligned}
\end{equation}
The saddle point equations yield,
\begin{equation}
    \begin{aligned}
        \hP[x,\hx] = {} & \rmi C \int  \calD y \calD \hy P(y,\hy) \overline{\exp\p{- \rmi \s{\alpha y \cdot \hx  + \alpha' x \cdot \hy }}} \\
        P[x,\hx] = {} & \sum_k p(k) \frac{k}{C} \frac{1}{\calA(k)}\exp\p{\rmi x \cdot \psi_k  + \rmi \hx \cdot \p{\frac{\dot x}{x} - [1 - x + h]}} \p{-\frac{i \hP(x, \hx)}{C}}^{k-1}
    \end{aligned}
\end{equation}

Which gives the final result
\begin{equation}
    \begin{aligned}
        \overline{Z[\bpsi]} \sim {} & \rme^{N S^\star[P]} \\
        S^\star[P] = {} & - \frac{C}{2} P \cdot \bW P + \sum_k p(k) \log \calA(k) 
    \end{aligned}
\end{equation}
where $P$ satisfies,
\begin{equation}
    P[x,\hx] = {} \sum_k p(k) \frac{k}{C} \frac{1}{\calA(k)}\exp\p{\rmi x \cdot \psi_k  + \rmi \hx \cdot \p{\frac{\dot x}{x} - [1 - x + h]}} \p{\bW P[x,\hx] }^{k-1}
\end{equation}
This is in general not possible to solve. It is a full complex distribution over paths $x(t),\hat{x}(t)$. Contrary to the mean-field case, it is not enough to know only the first two moments of the distribution to determine $P(x,\hat{x})$. For this reason we will follow the assumption of $C>>1$ and peform an expantion in $1/C$.

Since the degree distribution $p(k)$ will become ill-defined in the $C\to \infty$ limit, it is necessary to introduce a rescaled degree distribution, following \cite{metz2023Ising} we define
\begin{equation}
    \nu(g) = \lim_{C\to \infty} \sum_k p(k) \delta\p{g - \frac{k}{C}}
\end{equation}
We also look at the large $C$ expansion of $-i
\hP$, noticing that $\int \calD y \calD \hy P(y,\hy) = 1$
\begin{equation}
    \bW P \sim  1 - \rmi \frac{\mu}{C} (\expected{y} \cdot \hx  + \expected{\hy} \cdot x) - \frac{\sigma^2}{2 C} \hx \cdot \expected{ y y^\top }\hx - \frac{\sigma^2}{2 C} x \cdot \expected{ \hy \hy^\top }x - \frac{\gamma \sigma^2}{C} \hx \cdot \expected{ y \hy^\top } x
\end{equation}
Plugging this expansion in into the saddle point equation we get
\begin{equation}
    \begin{aligned}
        P[x,\hx] =  {} & \int \diff g \, \nu(g) g \frac{1}{\calA(g)} \exp\p{\rmi x \cdot \psi_g  + \rmi \hx \cdot \p{\frac{\dot x}{x} - [1 - x + h]} - g q[x,\hx]}\\
        q[x,\hx] = {} &  \rmi \mu (\expected{y} \cdot \hx  + \expected{\hy} \cdot x) + \frac{\sigma^2}{2 } \hx \cdot \expected{ y y^\top }\hx +\frac{\sigma^2}{2 } x \cdot \expected{ \hy \hy^\top }x +\gamma \sigma^2 \hx \cdot \expected{ y \hy^\top } x
    \end{aligned}
    \label{eq:saddle-g}
\end{equation}
where we have defined $\expected{\bullet} = \int \calD x \calD \hx \bullet P[x,\hx] $.
We now name the kernels, 
\begin{equation}
    \begin{aligned}
        \bC = {} & \expected{x x^\top} \\
        \rmi \bG = {} &  \expected{x \hx^\top} \\
        \bL = {} & \expected{\hx \hx^\top}
    \end{aligned}
\end{equation}

As an ansatz for the solution, we set $\bL = 0$, $\expected{\hx} = 0$, and $\psi_g =0$. Making the observation that under these assumptions we have that $\calA(g) = 1$, we can integrate over $\hx$ and then rewrite \eqref{eq:saddle-g} as,
\begin{equation}
    \begin{aligned}
        P[x] & = \int \rmd g \, \nu (g) g \, \calD \eta  \, \calP [\eta] P_{g,\eta}[x]\\
        P_{g,\eta}[x] & = \expected{\delta\s{\frac{\dot{x}}{x} - (1 - x + g \mu M 
 + \sqrt{g} \sigma \eta + g \gamma \sigma^2 \bG x)}}_{x_0}\\
        \calP [\eta]  & \propto \exp\p{- \frac{1}{2} \eta \cdot \bC^{-1} \eta}\\
        M(t) & = \expected{x(t)}_P\\
        C(t,s) & = \expected{x(t)x(s)}_P\\
        G(t,s) & = \int \rmd g \, \nu(g) \, g \, \calD \eta \, \calP [\eta] \expected{\frac{1}{\sqrt{g}\sigma }\frac{\delta x_g(t)}{\delta \eta(s)}}_{P_{g,\eta}}
    \end{aligned}
    \label{eq:cavity}
\end{equation}
where
\begin{equation}
    \begin{aligned}
        \bG x  & = \int_{0}^t \diff s \, G(t,s) x(s)\\
        \eta \cdot \bC^{-1} \eta & = \iint \diff t \diff s\, \eta(t) C^{-1}(t,s) \eta(s)
    \end{aligned}
\end{equation}
This clearly has a simple probabilistic interpretation. These functions $M$, $C$, and $G$, have to be determined self consistently from a dynamical process different from the original one, the \emph{cavity} one. This is almost the same except that the degree distribution is substituted by $\nu(g)g$ instead of $\nu(g)$. Once these functions are found, then one can find the statistics of the original problem following, 
\begin{equation}
    \begin{aligned}
        Q[x] & = \int \rmd g \, \nu (g) \, \calD \eta \calP [\eta] \, P_{g,\eta}[x]\\
        \calP [\eta]  & \propto \exp\p{- \frac{1}{2} \eta \cdot \bC^{-1} \eta}
    \end{aligned}
\end{equation}

\section{Numerical simulations}

All simulations where done in python using the package NetworkX, \cite{networkx} to generate the random graphs and the ordinary differential equation integrator of SciPy package \texttt{odeint}, \cite{scipy}. Initial conditions were always taken as a truncated Gaussian with $\mu_0 = 1$ and $\sigma_0 = 0.1$. Code is available upon request.

\end{document}